\documentclass[review]{elsarticle}
\usepackage[left=2cm,right=2cm,top=2cm,bottom=2cm]{geometry}
\usepackage{lineno,hyperref}
\modulolinenumbers[5]

\journal{Journal of \LaTeX\ Templates}

\usepackage{bm}
\usepackage[lofdepth,lotdepth]{subfig}
\usepackage{placeins}
\usepackage{amsmath}
\usepackage{amssymb}
\newcommand*{\norme}[1]{\left\lVert{#1}\right\rVert} 
\usepackage[]{algorithm2e}
\usepackage{longtable}
\usepackage{changepage}

\usepackage{array,multirow,makecell}
\setcellgapes{1pt}
\makegapedcells
\newcolumntype{R}[1]{>{\raggedleft\arraybackslash }b{#1}}
\newcolumntype{L}[1]{>{\raggedright\arraybackslash }b{#1}}
\newcolumntype{C}[1]{>{\centering\arraybackslash }b{#1}}

\begin{document}

\begin{frontmatter}

\title{Convergence issues in derivatives of Monte Carlo null-collision integral formulations: a solution}

%%% Group authors per affiliation:
\author[1]{J-M. Tregan}
\cortext[]{Corresponding author}
\ead{tregan@laplace.univ-tlse.fr}

\author[1]{S. Blanco}
\author[3]{J. Dauchet}
\author[2]{M. El Hafi}
\author[1]{R. Fournier}
\author[2]{L. Ibarrart}
\author[4]{P. Lapeyre}
\author[5,6]{N. Villefranque}

\address[1]{LAPLACE, UMR 5213 - Universit{\'{e}} Paul Sabatier, 118, Route de Narbonne - 31062 Toulouse Cedex, France}
\address[2]{Laboratoire RAPSODEE - UMR 5302 - ENSTIMAC - Campus Jarlard - 81013 Albi CT Cedex 09, France}
\address[3]{Universit\'e Clermont Auvergne, CNRS, SIGMA Clermont, Institut Pascal,
F-63000 Clermont-Ferrand, France}
\address[4]{PROMES - UPR CNRS 8521 - 7, rue du Four Solaire, 66120 Font Romeu Odeillo, France}
\address[5]{Centre National de Recherches M\'et\'eorologiques (CNRM), UMR 3589 CNRS, M\'et\'eo France, Toulouse}
\address[6]{Laboratoire Plasma et Conversion d'\'Energie (LAPLACE), UMR 5213 CNRS, Universit\'e Toulouse III}
\begin{abstract}
When a Monte Carlo algorithm is used to evaluate a physical observable $A$, it is possible to slightly modify the algorithm so that it evaluates simultaneously $A$ and the derivatives $\partial_\varsigma A$ of $A$ with respect to each problem-parameter $\varsigma$. The principle is the following: Monte Carlo considers $A$ as the expectation of a random variable, this expectation is an integral, this integral can be derivated as function of the problem-parameter to give a new integral, and this new integral can in turn be evaluated using Monte Carlo. The two Monte Carlo computations (of $A$ and $\partial_\varsigma A$) are simultaneous when they make use of the same random samples, i.e. when the two integrals have the exact same structure. It was proven theoretically that this was always possible, but nothing insures that the two estimators have the same convergence properties: even when a large enough sample-size is used so that $A$ is evaluated very accurately, the evaluation of $\partial_\varsigma A$ using the same sample can remain inaccurate. We discuss here such a pathological example: null-collision algorithms are very successful when dealing with radiative transfer in heterogeneous media, but they are sources of convergence difficulties as soon as sensitivity-evaluations are considered. We analyse theoretically these convergence difficulties and propose an alternative solution. 
\end{abstract}
%!!!!!!!!!!!!!!!!!!!!!!!!!!!!!!!!!!!!!!!!!!!!!!!!!!
\begin{keyword}
\texttt{Monte Carlo method} \sep \texttt{Direct derivatives} \sep \texttt{Null-collision algorithm} \sep \texttt{Sensitivity} \sep \texttt{Integral formulation}
\MSC[2010] 00-01\sep  99-00
\end{keyword}
%!!!!!!!!!!!!!!!!!!!!!!!!!!!!!!!!!!!!!!!!!!!!!!!!!!
\end{frontmatter}
%\linenumbers

\section{Introduction}
\label{sec:introduction}

When numerically simulating linear-transport physics using Monte Carlo algorithms, one of the most recurrent difficulties is the handling of highly non-homogenous or fast-variating media. This difficulty was encountered since the beginning of neutron-transport and plasma-physics modelling. But a quite elegant trick was soon identified as a way to bypass this difficulty~: virtual collisionners can be added where the true collisionners are scarce so that the total collisionner-density is homogeneous. Of course, in order to ensure that the physical problem is unchanged, when a particle interacts with a virtual collisionner, it simply continues its path as if no collision had occurred \citep{galtier2013,skullerud1968,woodcock1965,lin1978}. This is the meaning of the denomination \emph{null-collision algorithm} or \emph{fictitious-collision algorithm}\footnote{Similar keywords are \emph{pseudo-collision}, \emph{null-events}, \emph{fictitious-events}, \emph{null-collisions}, \emph{Woodcock tracking} and \emph{maximum cross-section}}. The first practical benefit is that the next collision event can be sampled as if the medium was homogenous. Then the choice is made to select a true-collision or a virtual-collision as function of their local respective-amounts and this is how the spatial information is recovered. But several other benefits were recently foreseen in \citep{galtier2013} and practically tested in \citep{galtier2016,galtier2017, kutz2017,novak2018,novak2014,raab2008, dauchet2018addressing,dauchet2013practice, charon2016, terree, 2019arXiv190201137V, dauchet2015calculation}, mainly for radiative-transfer applications. The main idea is that null-collision algorithms transform the non-linearity of Beer-extinction into a linear-recursive problem that Monte Carlo handles without approximation\citep{terree}. This was for instance used in \citep{galtier2016} to deal with absorption-spectra of molecular gases combining very numerous transitions: the summation over all transitions could be treated by the Monte Carlo algorithm itself, which was previously assumed impossible because this summation was inside the exponential of Beer-extinction. Similarly, the vanishing of the exponential allowed the extension of implicit Monte Carlo algorithms for inversion of absorption and scattering coefficients from intensity measurements \citep{galtier2017}. Outside radiative transfer, a very similar idea was used to solve Electromagnetic Maxwell equations for energy propagation in particle-ensembles of statistically-distributed shapes despite of the nonlinearity associated to the square of the electric field\citep{charon2016}. Again similar is the algorithm proposed in \citep{terree} solving Boltzmann equation for micro-fluidics applications despite of the nonlinearity of the collision operator.

Back to radiative-transfer applications, the ideas suggested in \citep{galtier2013} have motivated significant developments in the computer-graphics community for the cinema industry. Here the benefit of using null-collisions is that it extends to participating media (aerosols or clouds) the orthogonality between data-description and data-treatment that was at the heart of the most recent use of Monte Carlo for rendering complex scenes \citep{kutz2017,novak2014,novak2018,raab2008}. The algorithm is indeed processed without any knowledge of the exact spatial-information, and it is only when a collision occurs that access to the field is required: the interaction between the radiative-transfer algorithm and the field-data is strictly restricted to this very moment. This allows the implementation of numerous acceleration techniques with little changes by comparison with those developed for handling complex surfaces. One of these techniques consists in the setting of an acceleration grid, adjusting the amount of virtual collisionners so that the total collisionner-density is both homogeneous in part and close enough to the real density-field. This avoids the sampling of too many useless virtual-collisions. This is one of the starting points of the present paper: null-collision algorithms allow the use of any amount of virtual-collisionners but numerical efficiency justifies that one tries to reduce them to the minimum.

However, we show here that reducing the amount of virtual-collisionners to a minimum leads to convergence difficulties when evaluating sensitivities. Sensitivity evaluation is a very general feature of Monte Carlo techniques: when a Monte Carlo algorithm is used to evaluate a physical observable $A$, it is always possible to modify the algorithm in such a way that it evaluates both $A$ and the derivatives $\partial_\varsigma A$ of $A$ with respect to each problem-parameter $\varsigma$, and most commonly the corresponding implementation is quite straightforward \cite{delataillade2002,delatorre2014,roger2005,roger2004,blanco2006short,terree2015diffusion,Mikhailov}\footnote{This is not at all straightforward for domain-deformation sensitivities\citep{roger2005,roger2004}, but we here stick to pure parametric sensitivities}. But evaluating sensitivities using null-collision algorithms is pathological: the better we adjust the acceleration grid, the worse the statistical convergence rate. In Sec.~\ref{sec:convergence-difficulties} we will illustrate this pathological behaviour evaluating the transmissivity of a beam through a non-homogeneous column. Then we propose an alternative approach in Sec.~\ref{sec:alternative-approach} where the design of the sensitivity-evaluation algorithm starts from the standard integral solution of the Boltzmann equation, i.e. without virtual-collisionners. The resulting sampling requirements are then addressed with the null-collision approach viewed as a simple rejection-sampling approach. This introduces the cost of sampling an additional random variable, but at this cost the convergence difficulties vanish. We illustrate the numerical behaviour of this modified algorithm is Sec.~\ref{sec:simulation-examples} using a benchmark inspired of \citep{galtier2013}.

\section{Convergence difficulties when evaluating sensitivities}
\label{sec:convergence-difficulties}

In this section we design a Monte-Carlo algorithm using the standard null-collision approach for the evaluation of the distribution function $f$, i.e. the solution of Boltzmann equation, and apply the technique of \citep{roger2005,delataillade2002} for simultaneous evaluation of a sensitivity $\partial_\varsigma f $ with respect to $\varsigma $, where $\varsigma$ is a parameter appearing in the absorption and scattering coefficients. The Boltzmann equation that we use is linear with constant speed particles. It matches the monochromatic radiative-transfer equation exactly and all application examples will be restricted to radiative transfer. We choose to make use of the notation $f$ instead of the more radiative-transfer oriented notation $I=h\nu c f$ (the specific intensity) in order to simplify the access for readers of the plasma and neutronics communities. The monochromatic radiative-transfer equation becomes
\begin{equation}
\left\{
\begin{array}{rl}
&\partial_t f +c\bm{\vec{\omega}}.\bm{\vec{\nabla}} f= -(k_a + k_s) c f  +k_a c f^{eq} + \displaystyle \int_{4\pi} k_s c f' p_{\mathcal{S}}(-\bm{\vec{\omega}'}|-\bm{\vec{\omega}}) \, \mathrm{d}\omega', ~~~~\forall \bm{\vec{x}}\in\Omega, \forall \bm{\vec{\omega}}\in\mathbb{S}^2\\
& f(\bm{\vec{y}},\bm{\vec{\omega}}_+)=f_{\partial\Omega}(\bm{\vec{y}},\bm{\vec{\omega}}_+), ~~~~\forall \bm{\vec{y}}\in\partial\Omega,~ \forall \bm{\vec{\omega}}_+\in\mathbb{S}_+^2\\
& f(\bm{\vec{x}},\bm{\vec{\omega}},0,\varsigma)=f_0(\bm{\vec{x}},\bm{\vec{\omega}}), ~~~~\forall \bm{\vec{x}}\in\Omega, \forall \bm{\vec{\omega}}\in\mathbb{S}^2
\end{array}\right.
\label{boltz}
\end{equation}
where $f\equiv f(\bm{\vec{x}},\bm{\vec{\omega}},t,\varsigma)$ with $\bm{\vec{x}}$ the location, $\bm{\vec{\omega}}$ the propagation direction and $t$ the time. For incoming scattering in any direction $\bm{\vec{\omega}'}$ of the unit sphere $\mathbb{S}^2$, we write $f'\equiv f(\bm{\vec{x}},\bm{\vec{\omega}'},t,\varsigma)$ and $p_{\mathcal{S}}$ is the single scattering phase function, i.e. $p_{\mathcal{S}}(-\bm{\vec{\omega}'}|-\bm{\vec{\omega}})\mathrm{d}\omega'$ is the probability density that the scattering direction is $\bm{\vec{\omega}}$ for this incoming direction $\bm{\vec{\omega}'}$. The constant particle-speed is $c$ and  the coefficients $k_a\equiv k_a(\bm{\vec{x}},t,\varsigma)$, $k_s\equiv k_s(\bm{\vec{x}},t,\varsigma)$ and $k_e=k_a+k_s$ are the absorption coefficient, the scattering coefficient and the extinction coefficient respectively. $f^{eq}\equiv f^{eq}(\bm{\vec{x}},t)$ is the equilibrium distribution (following the Planck function). $\Omega$ is the geometrical domain and $\partial\Omega$ its boundary at which the distribution function $f_{\partial\Omega}$ is known for all locations $\bm{\vec{y}}$ and all directions $\bm{\vec{\omega}}_+$ of the incoming hemisphere $\mathbb{S}_+^2$. $f_0$ is the initial condition.

\begin{figure}[p]
\begin{center}
\includegraphics[page=1,clip,scale=0.9]{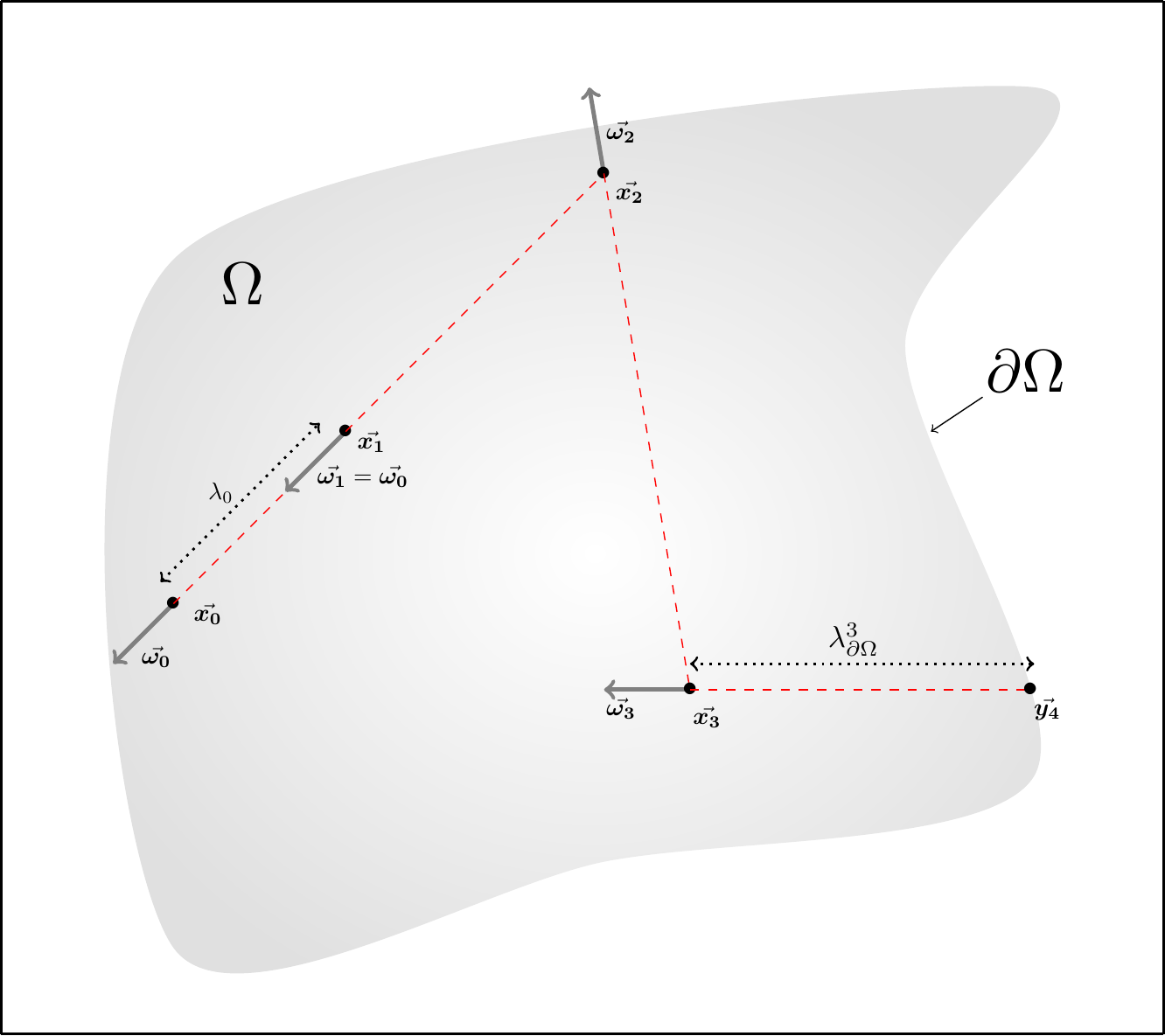}
\end{center}
\caption{
Example of a backward-sampled path obtained from Alg.~\ref{alg:Algof}, \ref{alg:AlgoDf1} or \ref{alg:AlgoDf2}. $\Omega$ is the geometrical domain and $\partial\Omega$ its boundary. $\vec{\bm{x_0}}$ and $\vec{\bm{\omega_0}}$ are the position and direction when starting the backward tracking (i.e. the end of the path) . This path is composed of one null-collision at location $\vec{\bm{x_1}}=\vec{\bm{x0}}-\lambda_0\vec{\bm{\omega_0}}$ (the initial direction $\vec{\bm{\omega_1}}$ is kept and therefore $\vec{\bm{\omega_0}}=\vec{\bm{\omega_1}}$), two scatterings at locations $\vec{\bm{x_2}}$ (from direction $\vec{\bm{\omega_2}}$ to direction $\vec{\bm{\omega_1}}$) and $\vec{\bm{x_3}}$ (from direction $\vec{\bm{\omega_3}}$ to direction $\vec{\bm{\omega_2}}$), and one boundary intersection at location $\vec{\bm{y_4}}=\vec{\bm{x_3}}-\lambda_{\partial\Omega}^3\vec{\bm{\omega_3}}$.}
\label{fig_Algof}
\end{figure}

\paragraph{Introducing null-collisions} In order to design a null collision algorithm (NCA)\citep{galtier2013} we add a field of virtual collisionners such that the total extinction coefficient is practicable, in the sense that we can sample the corresponding beer extinction:
\begin{adjustwidth}{-2cm}{0cm}
\begin{equation}
\left\{
\begin{array}{rl}
&\partial_t f +c\bm{\vec{\omega}}.\bm{\vec{\nabla}} f= -\hat{k} c f +k_a c f^{eq} + \displaystyle \int_{4\pi} k_s c f' p_{\mathcal{S}}(-\bm{\vec{\omega}'}|-\bm{\vec{\omega}}) \, \mathrm{d}\omega' + \displaystyle \int_{4\pi} k_n c f'\delta(\bm{\vec{\omega}}-\bm{\vec{\omega}'}) \, \mathrm{d}\omega', ~~~~\forall \bm{\vec{x}}\in\Omega, \forall \bm{\vec{\omega}}\in\mathbb{S}^2\\
& f(\bm{\vec{y}},\bm{\vec{\omega}}_+)=f_{\partial\Omega}(\bm{\vec{y}},\bm{\vec{\omega}}_+), ~~~~\forall \bm{\vec{y}}\in\partial\Omega, \forall \bm{\vec{\omega}}_+\in\mathbb{S}_+^2\\
& f(\bm{\vec{x}},\bm{\vec{\omega}},0,\varsigma)=f_0(\bm{\vec{x}},\bm{\vec{\omega}}), ~~~~\forall \bm{\vec{x}}\in\Omega, \forall \bm{\vec{\omega}}\in\mathbb{S}^2
\end{array}\right.
\label{boltz+khat}
\end{equation}
\end{adjustwidth}
where $k_n\equiv k_n(\bm{\vec{x}},t,\varsigma)$ is the null-collision coefficient, $\hat{k}=k_a+k_s+k_n$ is the total extinction-coefficient and $\delta$ is the Dirac distribution. Equation\eqref{boltz+khat} is strictly equivalent to Eq.\eqref{boltz} because of the Dirac distribution that insures $\int_{4\pi} k_n c f'\delta(\bm{\vec{\omega}}-\bm{\vec{\omega}'}) \, \mathrm{d}\omega' = k_n c f$.\\
When numerically adressing the solution $f(\bm{\vec{x}}_0,\bm{\vec{\omega}}_0)$ of this transport equation at $(\bm{\vec{x}}_0,\bm{\vec{\omega}}_0)$ (also solution of Eq.\eqref{boltz}) using the Monte Carlo method, one of the most standard approach consists in a simple statistical reading that allows to view $f(\bm{\vec{x}}_0,\bm{\vec{\omega}}_0)$ as an average over radiative paths that are tracked backward from the observation location $(\bm{\vec{x}}_0,\bm{\vec{\omega}}_0)$ to the sources\cite{galtier2013}. In this reading, the pure transport term $\partial_t f +c\bm{\vec{\omega}}.\bm{\vec{\nabla}} f$ corresponds to the spatial and temporal propagation of $f$ in direction $\bm{\vec{\omega}}$ at constant-speed $c$. The collisional term $-\hat{k} c f$ corresponds to either an absorption or a scattering event (including the null-collision events that are forward scattering events). When combining it with the transport term this leads to collision locations that are distributed exponentially along the line of sight (Beer law). Tracking the path backward, this means that the preceeding collision at $\bm{\vec{x}}_1$ is at a distance $\lambda_0$ that is a realisation of a random variable $\Lambda_0$ of probability density $p_{\Lambda_0}(\lambda_0) = \exp(-\hat{k} \lambda_0)$ (see Fig.~\ref{fig_Algof}). Once $\bm{\vec{x}}_1$ is sampled, the collision type is sampled in turn to decide wether an absorption, a true scattering or a null collision occurs. In the backward tracking picture, this corresponds respectively to the three remaining terms
\begin{itemize}
\item with $k_a c f^{eq}$ an absorption event is translated into thermal emission and the algorithm stops with the Monte Carlo weight $f^{eq}(\bm{\vec{x}}_1)$ (the source at $\bm{\vec{x}}_1$),
\item with $\int_{4\pi} k_s c f' p_{\mathcal{S}}(-\bm{\vec{\omega}'}|-\bm{\vec{\omega}}) \, \mathrm{d}\omega'$ a scattering event is translated into the sampling of a ``previous'' direction $\bm{\vec{\omega}}_1$ and the algorithm continues recursively as if evaluating $f(\bm{\vec{x}}_1,\bm{\vec{\omega}}_1)$,
\item with $\int_{4\pi} k_n c f'\delta(\bm{\vec{\omega}}-\bm{\vec{\omega}'}) \, \mathrm{d}\omega'$ and its Dirac function, a null collision event is translated into a pure forward scattering event, i.e. the ``previous'' direction $\bm{\vec{\omega}}_1$ is equal to $\bm{\vec{\omega}}_0$.
\end{itemize}
Of course the statistical translation includes the boundary conditions: when backward reaching the boundary at a location $\bm{\vec{x}}_i$ and direction $\bm{\vec{\omega}}_i$, the algorithm stops with the Monte Carlo weight $f(\bm{\vec{x}}_i, \bm{\vec{\omega}}_i)$ (the incoming source at the boundary). The corresponding Monte Carlo algorithm is detailed in Alg.~\ref{alg:Algof} and illustrated in Fig.~\ref{fig_Algof}.

\paragraph{Integral formulation} This null-collision algorithm belongs to the family of analog Monte Carlo algorithms, i.e. algorithms that can be designed without any formal development because they only numerically-implement the well established statistical pictures of radiation physics. However, in the present context it is very much useful to also choose a viewpoint under which the same algorithm appears as a statistical estimate of the integral solution of Eq.\eqref{boltz+khat}. For sake of clarity we only write this integral solution at the stationary limit:
\begin{equation}
\begin{array}{rl}
f(\bm{\vec{x}},\bm{\vec{\omega}},\varsigma)=&\exp\left(-\displaystyle \int_{0}^{\lambda_{\partial\Omega}} \hat{k}\left(\bm{\vec{\tilde{x}}}\right) \mathrm{d}\tilde{\lambda} \right)f_{\partial\Omega}(\bm{\vec{y}},\bm{\vec{\omega}})\\
+&\displaystyle \int_{0}^{\lambda_{\partial\Omega}}\exp\left(-\displaystyle \int_{0}^{\lambda} \hat{k}\left(\bm{\vec{\tilde{x}}}\right) \mathrm{d}\tilde{\lambda} \right)
\left(
\begin{array}{rl}
&k_a(\bm{\vec{x}'},\varsigma)f^{eq}(\bm{\vec{x}'})\\ 
+ &k_s(\bm{\vec{x}'},\varsigma) \displaystyle \int_{4\pi} p_S(-\bm{\vec{\omega}'}|-\bm{\vec{\omega}})\, \mathrm{d}\omega'f(\bm{\vec{x}'},\bm{\vec{\omega}'},\varsigma) \\ 
+ &k_n(\bm{\vec{x}'},\varsigma) f(\bm{\vec{x}'},\bm{\vec{\omega}},\varsigma)
\end{array}
\right)\mathrm{d}\lambda
\end{array}
\label{Sol_Fredholm_avec_khat}
\end{equation}
where $\bm{\vec{\tilde{x}}}=\bm{\vec{x}}-\tilde{\lambda}\bm{\vec{\omega}}$, $\bm{\vec{x}'}=\bm{\vec{x}}-\lambda\bm{\vec{\omega}}$, $\bm{\vec{y}}= \bm{\vec{x}}-\lambda_{\partial\Omega}\bm{\vec{\omega}}$, with $\lambda_{\partial\Omega}$ the distance to the first boundary-intersection starting at $\bm{\vec{x}}$ in the direction $-\bm{\vec{\omega}}$, i.e. $\lambda_{\partial\Omega}=\mathrm{min}\{\norme{\bm{\vec{x}}-\bm{z}}$;~$\bm{z}\in\mathrm{Vect}^-(\bm{\vec{x}},\bm{\vec{\omega}})\cap\partial\Omega \}$ where $\mathrm{Vect}^-(\bm{\vec{x}},\bm{\vec{\omega}})=\{\bm{\vec{x}}-\lambda'\bm{\vec{\omega}}$;~ $\lambda'\in\mathbb{R}_+\}$. This standard Fredholm equation, typical of the formal solution of linear-transport physics, can be transformed using the following property
$$\exp\left(-\displaystyle \int_{0}^{\lambda_{\partial\Omega}} \hat{k}\left(\bm{\vec{\tilde{x}}}\right) \mathrm{d}\tilde{\lambda} \right) = \displaystyle \int_{\lambda_{\partial\Omega}}^{+\infty}\hat{k}(\bm{\vec{x}'})\exp\left(-\displaystyle \int_{0}^{\lambda} \hat{k}\left(\bm{\vec{\tilde{x}}}\right) \mathrm{d}\tilde{\lambda} \right)\mathrm{d}\lambda$$
to give
\begin{equation}
\begin{array}{rl}
f(\bm{\vec{x}},\bm{\vec{\omega}},\varsigma)=&\displaystyle \int_{\lambda_{\partial\Omega}}^{+\infty}\hat{k}(\bm{\vec{x}'})\exp\left(-\displaystyle \int_{0}^{\lambda} \hat{k}\left(\bm{\vec{\tilde{x}}}\right) \mathrm{d}\tilde{\lambda} \right) f_{\partial\Omega}(\bm{\vec{y}},\bm{\vec{\omega}}) \mathrm{d}\lambda\\
+&\displaystyle \int_{0}^{\lambda_{\partial\Omega}}\hat{k}(\bm{\vec{x}'})\exp\left(-\displaystyle \int_{0}^{\lambda} \hat{k}\left(\bm{\vec{\tilde{x}}}\right) \mathrm{d}\tilde{\lambda} \right)
\left(
\begin{array}{rl}
&\frac{k_a(\bm{\vec{x}'},\varsigma)}{\hat{k}(\bm{\vec{x}'})}f^{eq}(\bm{\vec{x}'})\\ 
+ &\frac{k_s(\bm{\vec{x}'},\varsigma)}{\hat{k}(\bm{\vec{x}'})} \displaystyle \int_{4\pi} p_S(-\bm{\vec{\omega}'}|-\bm{\vec{\omega}})\, \mathrm{d}\omega'f(\bm{\vec{x}'},\bm{\vec{\omega}'},\varsigma) \\ 
+ &\frac{k_n(\bm{\vec{x}'},\varsigma)}{\hat{k}(\bm{\vec{x}'})} \displaystyle f(\bm{\vec{x}'},\bm{\vec{\omega}},\varsigma)
\end{array}
\right) \mathrm{d}\lambda
\end{array}
\label{Form_int_avec_khat0}
\end{equation}
Then
\begin{itemize}
\item $p_{\hat{\Lambda}}(\lambda)=\hat{k}(\bm{\vec{x}'})\exp\left(- \int_{0}^{\lambda} \hat{k}\left(\bm{\vec{\tilde{x}}}\right) \mathrm{d}\tilde{\lambda} \right)$ can be viewed as the probability density function of the free path $\hat{\Lambda}$ (the distance until next collision),
\item $P_A = \frac{k_a}{\hat{k}}$, $P_S = \frac{k_s}{\hat{k}}$ and $P_N = \frac{k_n}{\hat{k}}$ can be viewed as the probabilities that the collision is an absorption, a scattering event or a null-collision respectively,
\item and the two integrals over $[0,\lambda_{\partial\Omega}[$ and $[\lambda_{\partial\Omega},+\infty[$ can be gathered into a single integral over $[0,+\infty[$ using the  Heaviside function $\mathcal{H}$
\end{itemize}
to give
\begin{equation}
\begin{array}{rl}
f(\bm{\vec{x}},\bm{\vec{\omega}},\varsigma)=&\\
\displaystyle \int_{0}^{+\infty}p_{\hat{\Lambda}}(\lambda)\mathrm{d}\lambda&\left(
\begin{array}{rl}
&\mathcal{H}(\lambda-\lambda_{\partial\Omega})w_{\partial\Omega}\\
+&\mathcal{H}(\lambda_{\partial\Omega}-\lambda)
\left(
\begin{array}{rl}
&P_A(\bm{\vec{x}'},\varsigma)w_A\\ 
+ &P_S(\bm{\vec{x}'},\varsigma) \displaystyle \int_{4\pi} p_S(-\bm{\vec{\omega}'}|-\bm{\vec{\omega}})\, \mathrm{d}\omega'f(\bm{\vec{x}'},\bm{\vec{\omega}'},\varsigma) \\ 
+ &P_N(\bm{\vec{x}'},\varsigma) f(\bm{\vec{x}'},\bm{\vec{\omega}},\varsigma)
\end{array}
\right)
\end{array}
\right)
\end{array}
\label{Form_int_avec_khat1}
\end{equation}
with $w_{\partial\Omega}=f_{\partial\Omega}(\bm{\vec{y}},\bm{\vec{\omega}})$ and $w_A=f^{eq}(\bm{\vec{x}'})$. This last equation is the integral formulation that we needed in order to construct Alg.~\ref{alg:Algof} at the stationary limit: Alg.~\ref{alg:Algof} is indeed nothing more than the algorithmic-reading of Eq.~\ref{Form_int_avec_khat1} (and reciprocally Eq.~\ref{Form_int_avec_khat1} is nothing more than the integral translation of Alg.~\ref{alg:Algof}, \citep{delatorre2014}):
\begin{itemize}
\item $\int_{0}^{+\infty}p_{\hat{\Lambda}}(\lambda)\mathrm{d}\lambda$ stands for the sampling of the distance of the collision (according to the $\hat{k}$-field),
\item $\mathcal{H}(\lambda-\lambda_{\partial\Omega})$ stands for the case where the sampled collision is outside the boundary, then the algorithm stops at the boundary with the Monte Carlo weight $w_{\partial\Omega}$ (the value of $f$ corresponding to the incoming radiation),
\item $\mathcal{H}(\lambda_{\partial\Omega}-\lambda)$ stands for the case where the sampled collision is at a location $\bm{\vec{x}'}$ inside the volume, and then three collision types are possible:
\begin{itemize}
\item $P_A(\bm{\vec{x}'},\varsigma)$ stands for the case where the collision is an absorption, then the algorithm stops at the boundary with the Monte Carlo weight $w_A$ (the value of $f^{eq}$ at the collision location),
\item $P_S(\bm{\vec{x}'},\varsigma)$ stands for the case where the collision is a scattering event, then $\int_{4\pi} p_S(-\bm{\vec{\omega}'}|-\bm{\vec{\omega}})\, \mathrm{d}\omega'$ stands for the sampling of a new direction $\bm{\vec{\omega}'}$ according to the phase function and the algorithm continues recursively with the estimation of $f$ at $\bm{\vec{x}'}$ in direction $\bm{\vec{\omega}'}$,
\item $P_N(\bm{\vec{x}'},\varsigma)$ stands for the case where the collision is null, then the algorithm continues recursively with the estimation of $f$ at $\bm{\vec{x}'}$ in the unchanged direction $\bm{\vec{\omega}}$
\end{itemize}
\end{itemize}

\paragraph{Straightforward application of sensitivity-evaluation techniques} Now that we have constructed the integral formulation of Alg.~\ref{alg:Algof} we can apply the sensitivity-evaluation technique introduced in \citep{delataillade2002,roger2005,roger2004}. It consists in derivating Eq.~\ref{Form_int_avec_khat1} with respect to $\varsigma$ and multiplying and dividing by each of the probabilities and probability density functions that depend on $\varsigma$. This leads to an integral formulation of the sensitivity that has the very same structure as that of Eq.~\ref{Form_int_avec_khat1}:
\begin{equation}
\begin{array}{rl}
&\partial_{\varsigma}f(\bm{\vec{x}},\bm{\vec{\omega}},\varsigma) = \displaystyle \int_{0}^{+\infty}p_{\hat{\Lambda}}(\lambda)\mathrm{d}\lambda\\
&\left(
\begin{array}{rl}
&\mathcal{H}(\lambda-\lambda_{\partial\Omega})w^\varsigma_{\partial\Omega}\\
+&\mathcal{H}(\lambda_{\partial\Omega}-\lambda)
\left(
\begin{array}{rl}
&P_A(\bm{\vec{x}'},\varsigma)w^\varsigma_A\\ 
+ &P_S(\bm{\vec{x}'},\varsigma) \displaystyle \int_{4\pi} p_S(-\bm{\vec{\omega}'}|-\bm{\vec{\omega}})\mathrm{d}\omega' \left(
\begin{array}{rl}
&\frac{\partial_\varsigma k_s(\bm{\vec{x}'},\varsigma)}{k_s(\bm{\vec{x}'},\varsigma)}f(\bm{\vec{x}'},\bm{\vec{\omega}'},\varsigma)\\
+&\partial_\varsigma f(\bm{\vec{x}'},\bm{\vec{\omega}'},\varsigma)
\end{array}\right)\\ 
+ &P_N(\bm{\vec{x}'},\varsigma) \left(
\begin{array}{rl}
&\frac{\partial_\varsigma k_n(\bm{\vec{x}'},\varsigma)}{k_n(\bm{\vec{x}'},\varsigma)}f(\bm{\vec{x}'},\bm{\vec{\omega}},\varsigma)\\
+&\partial_\varsigma f(\bm{\vec{x}'},\bm{\vec{\omega}},\varsigma)
\end{array}\right)
\end{array}
\right)
\end{array}
\right)
\end{array}
\label{Form_int_s_avec_khat1}
\end{equation}
with $w^\varsigma_{\partial\Omega} = 0$ and $w^\varsigma_A = \frac{\partial_\varsigma k_a(\bm{\vec{x}'},\varsigma)}{k_a(\bm{\vec{x}'},\varsigma)}f^{eq}(\bm{\vec{x}'})$. Because of their identical structure, we can gather Eq.~\ref{Form_int_avec_khat1} and \ref{Form_int_s_avec_khat1} into one using the vectorial notation $\left\{ w ; w^\varsigma \right\}$:
\begin{adjustwidth}{-2cm}{0cm}
\begin{equation}
\begin{array}{rl}
&\left\{ f(\bm{\vec{x}},\bm{\vec{\omega}},\varsigma) ; \partial_{\varsigma}f(\bm{\vec{x}},\bm{\vec{\omega}},\varsigma) \right\}= \displaystyle \int_{0}^{+\infty}p_{\hat{\Lambda}}(\lambda)\mathrm{d}\lambda\\
&\left(
\begin{array}{rl}
&\mathcal{H}(\lambda-\lambda_{\partial\Omega})\left\{ w_{\partial\Omega} ; w^\varsigma_{\partial\Omega} \right\}\\
+&\mathcal{H}(\lambda_{\partial\Omega}-\lambda)
\left(
\begin{array}{rl}
&P_A(\bm{\vec{x}'},\varsigma)\left\{ w_A ; w^\varsigma_A \right\}\\ 
+ &P_S(\bm{\vec{x}'},\varsigma) \displaystyle \int_{4\pi} p_S(-\bm{\vec{\omega}'}|-\bm{\vec{\omega}})\mathrm{d}\omega' \left\{
f(\bm{\vec{x}'},\bm{\vec{\omega}'},\varsigma) ;
\left(
\begin{array}{rl}
&\frac{\partial_\varsigma k_s(\bm{\vec{x}'},\varsigma)}{k_s(\bm{\vec{x}'},\varsigma)}f(\bm{\vec{x}'},\bm{\vec{\omega}'},\varsigma)\\
+&\partial_\varsigma f(\bm{\vec{x}'},\bm{\vec{\omega}'},\varsigma)
\end{array}\right)\right\}\\ 
+ &P_N(\bm{\vec{x}'},\varsigma) \left\{
f(\bm{\vec{x}'},\bm{\vec{\omega}},\varsigma) ;
\left(
\begin{array}{rl}
&\frac{\partial_\varsigma k_n(\bm{\vec{x}'},\varsigma)}{k_n(\bm{\vec{x}'},\varsigma)}f(\bm{\vec{x}'},\bm{\vec{\omega}},\varsigma)\\
+&\partial_\varsigma f(\bm{\vec{x}'},\bm{\vec{\omega}},\varsigma)
\end{array}\right)\right\}
\end{array}
\right)
\end{array}
\right)
\end{array}
\label{Form_int_avec_khat2}
\end{equation}
\end{adjustwidth}
The algorithmic-reading of \ref{Form_int_avec_khat2} leads to Alg.~\ref{alg:AlgoDf1} that evaluates simultaneously $f$ and $\partial_{\varsigma}f$. The recursive nature of this algorithm comes from the fact that the final brackets in the scattering and null-collision terms contain $f$ and $\partial_\varsigma f$ at the same location in the same direction. The fact that their sensitivity part includes a summation is translated into an algorithm incrementing the Monte Carlo weight as explained in Appendix~\ref{appendix:weight}.

\begin{figure}[p]
\begin{center}
\includegraphics[scale=0.8]{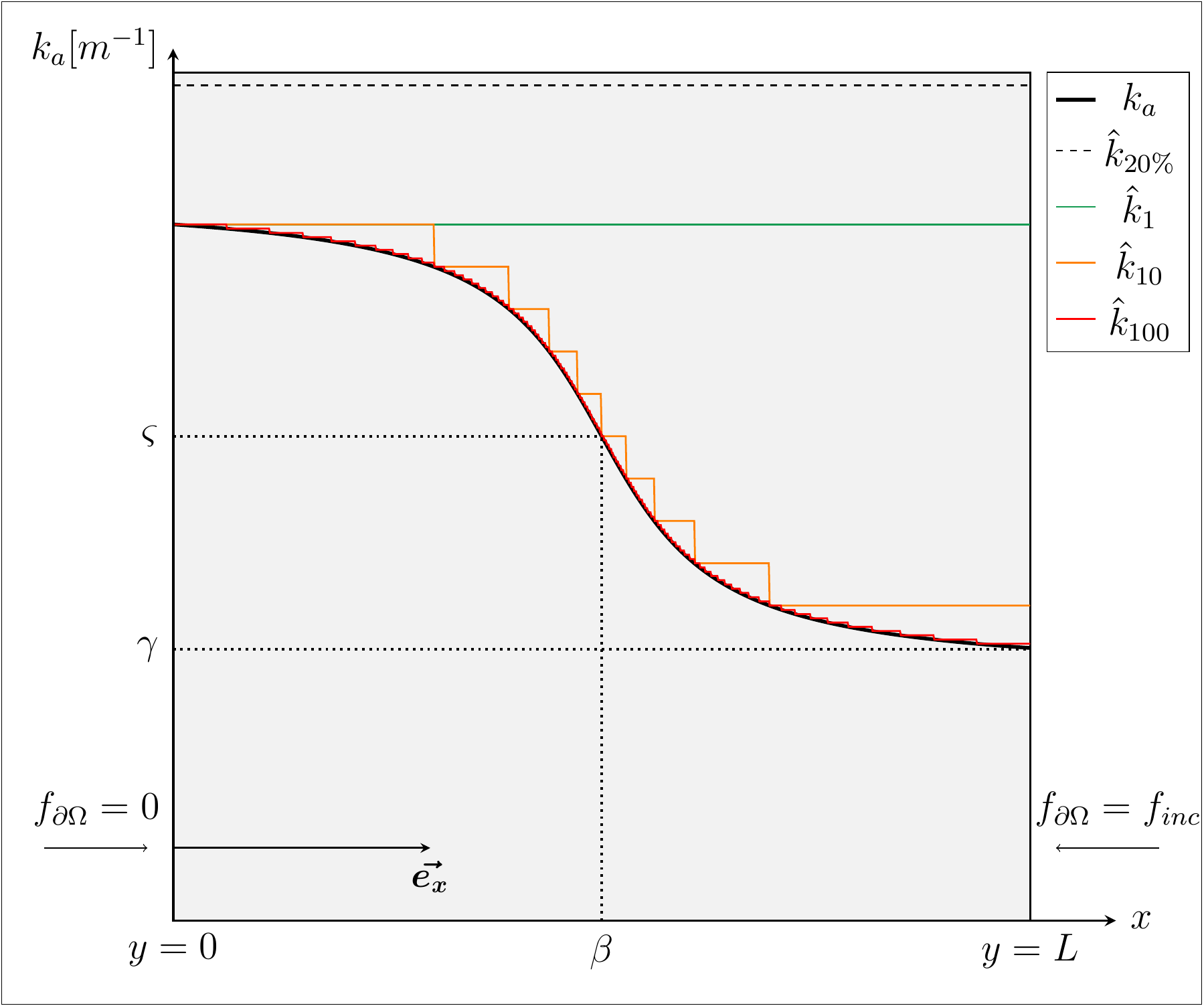}
\caption{
The \emph{heterogeneous-slab} configuration: a non diffusive heterogeneous column of length $L$. The absorption coefficient profile is $k_a(\bm{\vec{x}},\varsigma)=\left(\varsigma-\gamma \right)\frac{\mathrm{atan}(-\alpha\left(\bm{\vec{x}}-\beta\right)+\frac{\pi}{2})}{\pi/2}+\gamma$. Different $\hat{k}$ profiles are used, closer and closer to $k_e \equiv k_a$: $\hat{k}_{20\%}$ is equal to $1.2\mathrm{max}\left(k_e\right)$ and $\hat{k}_n$ is constant in pieces on the grid: $\hat{k}$ is exactly equal to the maximum $k_e$-value inside a mesh and the grid is constructed in such a way that the variations of $k_e$ are identical across each mesh). $\alpha, \beta$, $\varsigma$ and $\gamma$ are arbitrary parameters that allow us to choose the shape of the absorption coefficient profile, in particular, $\varsigma$ is the problem-parameter. The equilibrium distribution $f^{eq}$ is null (no emission) and the boundary conditions are $f_{\partial\Omega}(L,-\bm{\vec{e_x}})=f_{inc}$ and  $f_{\partial\Omega}(0,\bm{\vec{e_x}})=0$.
By choosing this $k_e$-profile, we can calculate the transmissivity and the sensibility analytically: $T=\mathrm{exp}\left(-\left(\varsigma-\gamma\right)K - \gamma L\right)$ and $\partial_\varsigma T= -K \mathrm{exp}\left(-\left(\varsigma-\gamma\right)K - \gamma L\right)$ where $K=\frac{1}{\varsigma/2}\left(\left(L-\beta\right)\mathrm{atan}\left(-\alpha L+\alpha\beta\right)+\beta\mathrm{atan}\left(\alpha\beta\right)-\frac{1}{2\alpha}\mathrm{log}\left(\frac{1+\left(\alpha\beta\right)^2}{1+\left(-\alpha L+\alpha\beta\right)^2}\right)\right)+ L$.}
\label{fig1}
\end{center}
\end{figure}

\paragraph{Simulation examples} At this stage, we designed a null-collision algorithm, constructed the corresponding integral formulation and applied the proposition of \citep{delataillade2002,roger2005,roger2004} in a straightforward manner so that the algorithm also evaluates sensitivities. We now test this simulation strategy by evaluating the transmissivity of a non-diffusive heterogeneous column and also evaluating the sensitivity of this transmissivity w.r.t. $\varsigma$, a parameter influencing the absorption coefficient. Hereafter this configuration is called \emph{heterogeneous-slab} (see Fig.~\ref{fig1}): $\Omega$ is a column of length $L$ with $\bm{\vec{e_x}}$ the normal incoming at location $y=0$. The equilibrium distribution is null (cold medium, $f^{eq}\equiv0$). The boundary conditions are $f^{\partial\Omega}(0,\bm{\vec{e_x}})=0$ and $f_{\partial\Omega}(L,-\bm{\vec{e_x}})=f_{inc}$. The absorption and scattering coefficients are $k_a(x,\varsigma)=\left(\varsigma-\gamma \right)\frac{\mathrm{atan}(-\alpha\left(x-\beta\right)+\frac{\pi}{2})}{\pi/2}+\gamma$ and $k_s\equiv0$. Alg~\ref{alg:AlgoDf1} is used to evaluate both $f(0,\bm{\vec{e_x}},\varsigma)$ and $\partial_{\varsigma}f(0,\bm{\vec{e_x}},\varsigma)$ that correspond to the transmissivity $T$ and its derivative $\partial_{\varsigma}T$ respectively: $T = f(0,\bm{\vec{e_x}},\varsigma)/f_{inc}$ and $\partial_{\varsigma}T = \partial_{\varsigma}f(0,\bm{\vec{e_x}},\varsigma)/f_{inc}$. We chose this particular profile of $k_a$, because it is possible to calculate $T$ and $\partial_{\varsigma}T$ analytically (see the caption of Fig.~\ref{fig1}). Example Monte Carlo results, using $N=10000$ samples, are compared to the analytical solution in Tables~\ref{table1} and \ref{table2}. The statistical uncertainty is noted $\sigma$ (the standard deviation of the Monte Carlo estimator). In Table~\ref{table2} we also provide the number of samples $N_{1\%}$ required to achieve a $1\%$ accuracy. The simulations were made using five different $\hat{k}$-profiles (each overestimating $k_a$ at all locations), with acceleration-grids, $\hat{k}$ being uniform within each mesh (see Fig.~\ref{fig1}):
\begin{itemize}
\item for $\hat{k}_{20\%}$ no grid is used: the profile of $\hat{k}$ is uniform, equal to $1.2$ time the maximum $k_a$-value. 
\item for $\hat{k}_1$ no grid is used: the profile of $\hat{k}$ is again uniform, exactly equal to the maximum $k_a$-value. 
\item for $\hat{k}_{10}$ the grid is constructed in such a way that across each mesh the variations of $k_a$ are $1/10$ of the maximum $k_a$-value, and the profile of $\hat{k}$ is uniform within each mesh, exactly equal to the maximum $k_a$-value inside the mesh.
\item for $\hat{k}_{100}$ and $\hat{k}_{1000}$ the grid is constructed the same way with $1/100$ and $1/1000$ variation respectively.
\end{itemize}
The transmissivity results of Table~\ref{table1} confirm that the estimation of $T$ is insensitive to the adjustment of the $\hat{k}$-field (only the computation time is affected). But the sensitivity results of Table~\ref{table2} clearly indicate the opposite: the statistical convergence is worse when $\hat{k}$ is close to $k$ and the number of samples required to reach a given accuracy level can be risen up to infinity when matching $\hat{k}$ to $k$ exactly. This is the pathological behavior that we announced in introduction: sensitivities cannot be evaluated accurately when using acceleration grids reducing the number of virtual collisions. 

\begin{figure}[h]
  \begin{center}
    \subfloat[Alg.~\ref{alg:Algof}]{
      \includegraphics[page=1,clip,scale=0.65]{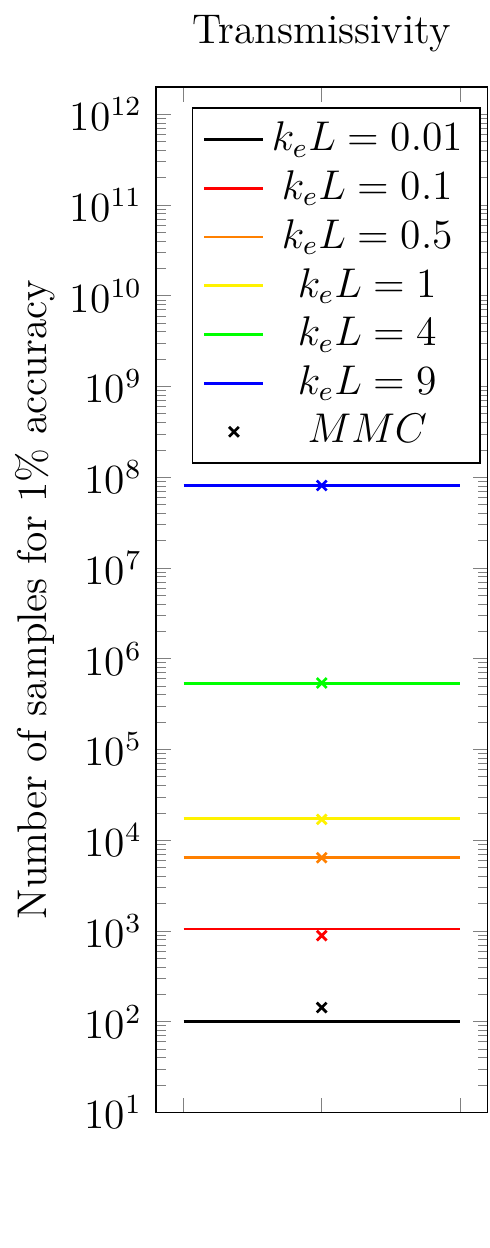}
      \label{hom_algo1}
                         }
    \subfloat[Alg.~\ref{alg:AlgoDf1} without null-collision]{
      \includegraphics[page=1,clip,scale=0.65]{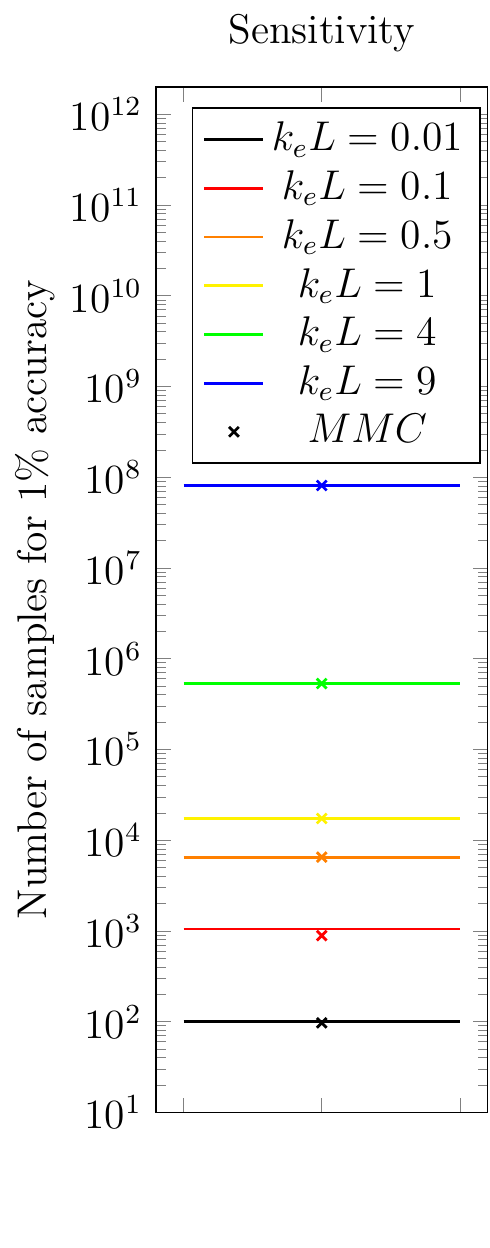}
      \label{hom_algo2_0}
                         }
    \subfloat[Alg.~\ref{alg:AlgoDf1} with null-collision: the standard approach]{
      \includegraphics[page=1,clip,scale=0.65]{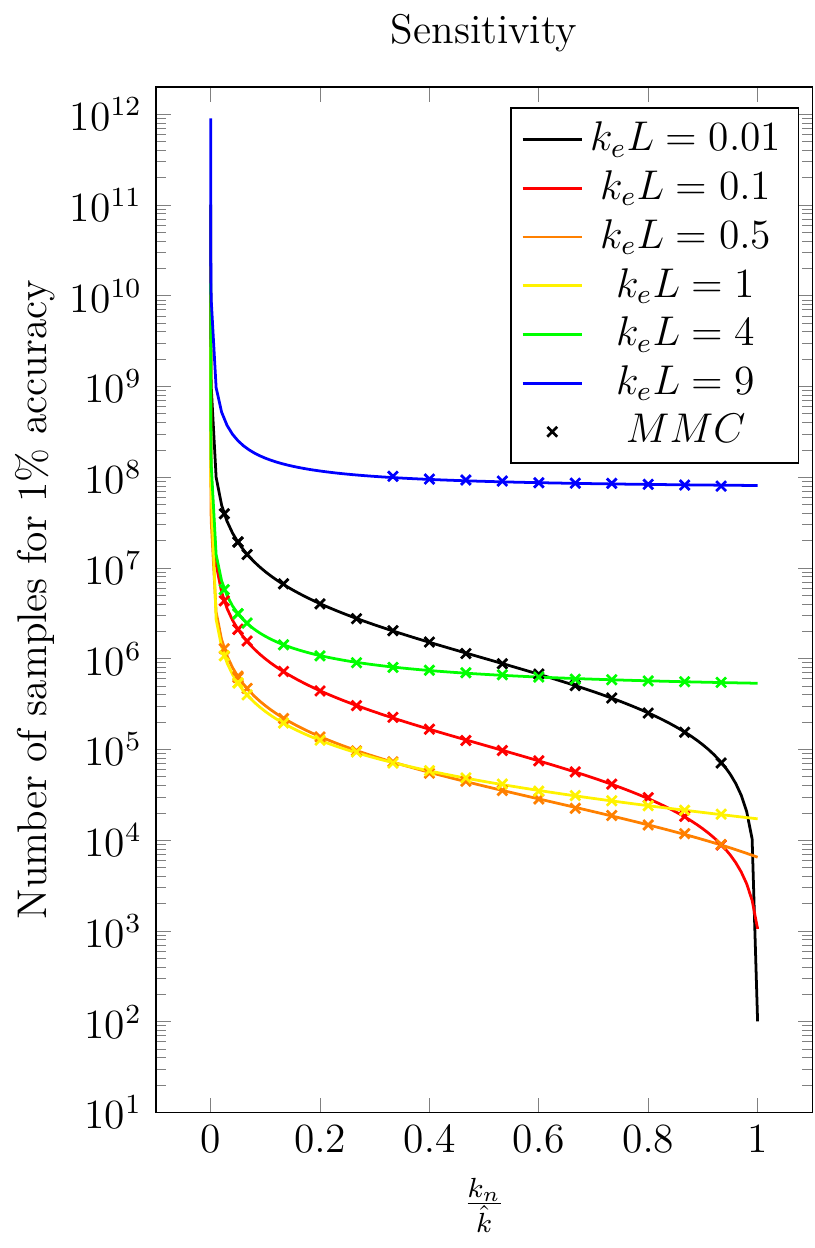}
      \label{hom_algo2_1}
                         }
      \subfloat[Alg.~\ref{alg:AlgoDf2} with null-collision: the alternative approach]{
      \includegraphics[page=1,clip,scale=0.65]{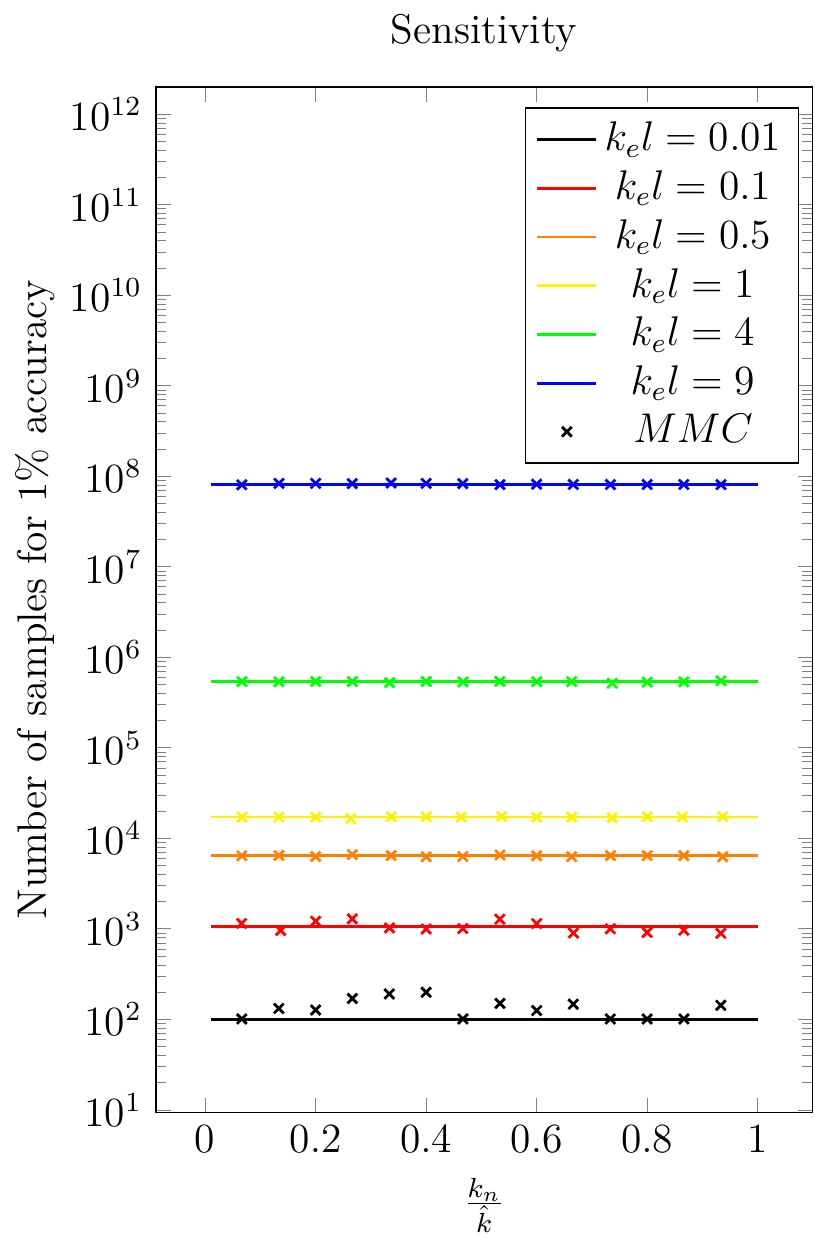}
      \label{hom_algo3}
                         }
\caption{
Statistical convergence for the \emph{homogeneous-slab} configuration: number of samples required to reach a $1\%$ accuracy (N$1\%$) when evaluating the slab transmissivity or its sensitivity as function of the ratio $\frac{k_n}{k}$ (i.e. the amount of null-collisions), for optical-thicknesses $k_e L \equiv k_a L$ ranging from $0.01$ to $9$. In this particular case the variances of Alg.~\ref{alg:Algof},  \ref{alg:AlgoDf1} and \ref{alg:AlgoDf2} are known analytically (see Appendix~\ref{appendix:homogeneous-slab}) and therefore N$1\%$ is also known analytically. It is displayed using solid lines. Also displayed are some example Monte Carlo simulations, only confirming that the analytical prediction is correct. These theoretical analysis of statistical convergence are provided for the the following four algorithms:
(a.) transmissivity estimation, with null-collision, using Alg.~\ref{alg:Algof} (the standard null-collision algorithm described in Section.~\ref{sec:convergence-difficulties});
(b.) sensitivity estimation, without null-collision, using Alg.~\ref{alg:AlgoDf1} (the standard sensitivity evaluation algorithm described in Section.~\ref{sec:convergence-difficulties}), here with $k_n\equiv0$ which is possible in this particular case because $k_e$ is homogeneous;
(c.) sensitivity estimation, with null-collision, using the same standard approach as in (b.) (i.e. Alg.~\ref{alg:AlgoDf1}, but now with $k_n \ne 0$);
(d.) sensitivity estimation, with null-collision, using the alternative approach (i.e. algorithm Alg.~\ref{alg:AlgoDf2} described in Section.~\ref{sec:alternative-approach}). 
}
\label{homogeneous_slab}
\end{center}
\end{figure}

\paragraph{The variance of the sensitivity estimate} For a better understanding of this behavior, we studied a \emph{homogeneous-slab} for which the variance of the Monte Carlo estimate can be calculated analytically. This case is identical to the previous one (transmissivity of a purely absorbing column) but now $k=k_a$ is uniform: $k_a(\varsigma) \equiv \varsigma$, $T = exp(-\varsigma L)$ and $\partial_{\varsigma}T = -L T$. Of course, there is no need to make use of a null-collision algorithm as soon as $k$ is uniform. We only do it for theoretical reasons (with $\hat{k}>k$ uniform). This allows us to fully identify the reasons why the variance of the sensitivity estimate rises when reducing $k_n = \hat{k}-k$. This may sound trivial as soon as when encountering a null-collision event, the Monte Carlo weight of the sensitivity algorithm includes a factor $\frac{\partial_\varsigma k_n(\bm{\vec{x}'},\varsigma)}{k_n(\bm{\vec{x}'},\varsigma)}=1/k_n$ (see Eq.~\ref{Form_int_avec_khat2}), but reducing $k_n$ also reduces the number of such null-collision occurrences. This may lead to a compensation, maintaining the variance at a finite value. The developments of appendix~\ref{appendix_anal_var_algo_2} indicate the opposite: the statistical uncertainty is indeed
\begin{equation}
  \sigma_{\partial_\varsigma T} = \frac{\sqrt{L^2\mathrm{e}^{-k_a L} \Big(\frac{k_n +1/L}{k_n}\Big)- L^2 \mathrm{e}^{-2 k_a L}}}{\sqrt{N}}
  \label{eq:explosion}
\end{equation}
Figure~\ref{homogeneous_slab} illustrates the meaning of this dependance of $\sigma_{\partial_\varsigma T}$ with the problem parameters. In this idealised case, looking at the behavior of such an algorithm applying sensitivity-evaluation techniques in a straightforward manner, the difficulty is well identified: when $\frac{k_n}{\hat{k}}$ approaches zero, the number of samples required for a $1\%$ accurate evaluation of the sensitivity tends to infinity (see Fig.~\ref{hom_algo2_1}). This figure also displays the behavior of an algorithm implementing the very same sensitivity-evaluation technique, but without the use of null-collisions (which is possible here in this idealised uniform case). Without null-collisions, the relative value of the standard deviation of the sensitivity-estimate (Fig.~\ref{hom_algo2_0}) is identical to that of the main quantity (the transmissivity-estimate, Fig.~\ref{hom_algo1}). This is an ideal behaviour: the sensitivity is estimated with the same relative accuracy as that of the main quantity. Altogether in this simple example, we see that evaluating sensitivities can be perfectly costless \emph{before} using null-collisions and may become pathological \emph{when null-collisions are introduced}.

Note that in the general case, even without null-collisions, evaluating sensitivities can be truely difficult. Understanding the relative variance of sensitivity estimates and comparing them to the relative variance of the algorithm estimating the main quantity was indeed one of the main concerns of the initial work of De~Lataillade\citep{delataillade2002}. Essentially, serious difficulties arise as soon as the scattering optical-thickness is high. The objective of the present paper is not at all to address this specific issue: at the end of the following section, when an alternative solution will be proposed for evaluating sensitivities in null collisions algorithms, the problems associated to highly scattering media will remain unsolved.

\section{An alternative approach}
\label{sec:alternative-approach}

The preceding section identifies convergence difficulties when evaluating sensitivities using null-collisions. Theses difficulties are not associated to the standard sensitivity-evaluation algorithm itself: considering slab transmission, we have seen that when we do not make use of null-collisions, the sensitivity-evaluation algorithm converges as well as the algorithm evaluating the main quantity. So the observed difficulties are only the consequences of introducing virtual-collisionners. At this stage, null-collision algorithms appear therefore as perfect tools for handling heterogeneous fields, but are incompatible with the simultaneous evaluation of sensitivities.

We have seen that this problem is related to the term $\frac{1}{k_n}$ appearing in the Monte Carlo weight of the sensitivity algorithm. At which stage did this term appear and can we bypass this step? Clearly, $\frac{1}{k_n}$ appeared when derivating with $\varsigma$ the null-collision probability $P_N(\varsigma) = 1-P_A(\varsigma)-P_S(\varsigma)$, with $P_A(\varsigma) = k_a(\varsigma)/\hat{k}$ and $P_S = k_s(\varsigma)/\hat{k}$. A first way to suppress this $\frac{1}{k_n}$ term consists in making $\hat{k}$ dependent on $\varsigma$. This is always possible because $\hat{k}$ is a free parameter and we can therefore adjust it to the variation of $k_a(\varsigma)+k_s(\varsigma)$ so that $P_N$ does not depend on $\varsigma$ anymore. We first tested this solution and it proved itself already quite practical: the corresponding details are provided in Appendix~\ref{appendix-k-chapeau-variable}. But we finally retained another algorithm, starting from the integral solution of the original Boltzmann Eq.~\ref{boltz}, i.e. prior to the introduction of virtual-collisionners. The idea consists in first designing an algorithm evaluating simultaneously $f$ and $\partial_{\varsigma}f$ \emph{as if the heterogeneity of the field could be handled without difficulty} and only introduce null-collisions in a second stage. For this, we can simply rewrite Eq.~\ref{Form_int_avec_khat1} with $k_n = 0$ (no virtual collisionners):
\begin{equation}
\begin{array}{rl}
f(\bm{\vec{x}},\bm{\vec{\omega}},\varsigma) &= \displaystyle \int_{0}^{+\infty}p_{\Lambda}(\lambda)\mathrm{d}\lambda\\
&\left(
\begin{array}{rl}
&\mathcal{H}(\lambda-\lambda_{\partial\Omega})w_{\partial\Omega}\\
+&\mathcal{H}(\lambda_{\partial\Omega}-\lambda)
\left(
\begin{array}{rl}
&P_A(\bm{\vec{x}'},\varsigma)w_A\\ 
+ &P_S(\bm{\vec{x}'},\varsigma) \displaystyle \int_{4\pi} p_S(-\bm{\vec{\omega}'}|-\bm{\vec{\omega}})\, \mathrm{d}\omega'f(\bm{\vec{x}'},\bm{\vec{\omega}'},\varsigma)
\end{array}
\right)
\end{array}
\right)
\end{array}
\label{Form_int_sans_khat1}
\end{equation}
The only differences with Eq.~\ref{Form_int_avec_khat1} are that
\begin{itemize}
\item $P_N = 0$,
\item the random variable $\hat{\Lambda}$ of probability density $p_{\hat{\Lambda}}(\lambda)=\hat{k}(\bm{\vec{x}'})\exp\left(-\int_{0}^{\lambda} \hat{k}\left(\bm{\vec{\tilde{x}}}\right) \mathrm{d}\tilde{\lambda} \right)$ (the free path in the $\hat{k}$-field) is replaced with the random variable $\Lambda$ of probability density $p_{\Lambda}(\lambda)=k_e(\bm{\vec{x}'},\varsigma)\exp\left(-\int_{0}^{\lambda} k_e(\bm{\vec{\tilde{x}}},\varsigma) \mathrm{d}\tilde{\lambda} \right)$ (the free path in the original $k_e$-field).
\end{itemize}
This equation can then be derivated with respect to $\varsigma$ and multiplied/divided by each of the probabilities and probability density functions that depend on $\varsigma$ (exactly the same way Eq.~\ref{Form_int_avec_khat2} was constructed from Eq.~\ref{Form_int_avec_khat1}) to give
\begin{equation}
\begin{array}{rl}
&\partial_{\varsigma}f(\bm{\vec{x}},\bm{\vec{\omega}},\varsigma) = \displaystyle \int_{0}^{+\infty}p_{\Lambda}(\lambda)\mathrm{d}\lambda\\
&\left(
\begin{array}{rl}
&\mathcal{H}(\lambda-\lambda_{\partial\Omega})
\left(
- w_{\partial\Omega} \displaystyle \int_{0}^{\lambda_{\partial\Omega}} \partial_\varsigma k_e(\bm{x_l},\varsigma) \mathrm{d}l 
+ w^\varsigma_{\partial\Omega}
\right)\\
+&\mathcal{H}(\lambda_{\partial\Omega}-\lambda)
\left(
\begin{array}{rl}
&P_A(\bm{\vec{x}'},\varsigma) \left( - w_A \displaystyle \int_{0}^{\lambda} \partial_\varsigma k_e(\bm{x_l},\varsigma) \mathrm{d}l + w^\varsigma_A \right)\\ 
+ &P_S(\bm{\vec{x}'},\varsigma) \displaystyle \int_{4\pi} p_S(-\bm{\vec{\omega}'}|-\bm{\vec{\omega}})\mathrm{d}\omega'\\
&\hspace{2cm}\left(
\begin{array}{rl}
-&\displaystyle \int_{0}^{\lambda} \partial_\varsigma k_e(\bm{x_l},\varsigma) \mathrm{d}l f(\bm{\vec{x}'},\bm{\vec{\omega}'},\varsigma)\\
+&\frac{\partial_\varsigma k_s(\bm{\vec{x}'},\varsigma)}{k_s(\bm{\vec{x}'},\varsigma)}f(\bm{\vec{x}'},\bm{\vec{\omega}'},\varsigma)\\
+& \partial_\varsigma f(\bm{\vec{x}'},\bm{\vec{\omega}'},\varsigma)
\end{array}
\right) 
\end{array}
\right)
\end{array}
\right)
\end{array}
\label{Form_int_sans_khat2}
\end{equation}
A main point of the present paper is that this integral equation, although it was derived the same way as Eq.~\ref{Form_int_avec_khat2}, cannot be interpreted in algorithmic terms: the integral pattern $\int \partial_\varsigma k_e \mathrm{d}l$ is not yet transformed into a statistical expectation. An additional random generation will be required. At this stage let us introduce an arbitrary random variable $L$ of probability density function $p_L$ and write $\int \partial_\varsigma k_e \mathrm{d}l = \int  p_L(l) \mathrm{d}l \frac{\partial_\varsigma k_e}{p_L(l)}$. Reporting this into Eq.~\ref{Form_int_sans_khat2} and using $\int p_L(l) \mathrm{d}l = 1$ leads to
\begin{equation}
\begin{array}{rl}
&\partial_{\varsigma}f(\bm{\vec{x}},\bm{\vec{\omega}},\varsigma) = \displaystyle \int_{0}^{+\infty}p_{\Lambda}(\lambda)\mathrm{d}\lambda\\
&\left(
\begin{array}{rl}
&\mathcal{H}(\lambda-\lambda_{\partial\Omega})\displaystyle \int_{0}^{\lambda_{\partial\Omega}} p_L(l|\lambda_{\partial\Omega})  \mathrm{d}l 
\left(
- w_{\partial\Omega}\frac{  \partial_\varsigma k_e(\bm{x_l},\varsigma)}{p_L(l|\lambda_{\partial\Omega})}
+ w^\varsigma_{\partial\Omega}
\right)\\
+&\mathcal{H}(\lambda_{\partial\Omega}-\lambda)\displaystyle \int_{0}^{\lambda} p_L(l|\lambda) \mathrm{d}l
\left(
\begin{array}{rl}
&P_A(\bm{\vec{x}'},\varsigma) \left( - w_A\frac{  \partial_\varsigma k_e(\bm{x_l},\varsigma)}{p_L(l|\lambda)}  + w^\varsigma_A \right)\\ 
+ &P_S(\bm{\vec{x}'},\varsigma) \displaystyle \int_{4\pi} p_S(-\bm{\vec{\omega}'}|-\bm{\vec{\omega}})\mathrm{d}\omega'\\
&\hspace{2cm}\begin{array}{rl} 
&\left(
\begin{array}{rl}
-& \frac{\partial_\varsigma k_e(\bm{x_l},\varsigma)}{p_L(l|\lambda)}f(\bm{\vec{x}'},\bm{\vec{\omega}'},\varsigma)\\
+&\frac{\partial_\varsigma k_s(\bm{\vec{x}'},\varsigma)}{k_s(\bm{\vec{x}'},\varsigma)}f(\bm{\vec{x}'},\bm{\vec{\omega}'},\varsigma)\\
+& \partial_\varsigma f(\bm{\vec{x}'},\bm{\vec{\omega}'},\varsigma)
\end{array}
\right)
\end{array} 
\end{array}
\right)
\end{array}
\right)
\end{array}
\label{Form_int_sans_khat3}
\end{equation}
At this stage, null-collisions have not been introduced. Therefore, the algorithmic reading of Eq.~\ref{Form_int_sans_khat3} would not be practical as soon as the $k_e$-field is heterogeneous: the difficulty would come from the sampling of $\Lambda$. The objective of introducing null collisions will therefore be to replace $\Lambda$ with another path-length $\hat{\Lambda}$, shorter in average but easy to sample, and compensate the too many collisions by the fact that some of them are null. However, this not as trivial as in the algorithm for the main quantity because of the new random variable $L$ that we needed to introduce when transforming Eq.~\ref{Form_int_sans_khat2} into Eq.~\ref{Form_int_sans_khat3} (transforming it into an expectation). Indeed $\partial_\varsigma k_e$ needs to be integrated along the whole path, now including null collisions. In statistical terms, this means that the recursivity of the path-sampling algorithm is only insured if the Monte Carlo weight associated to null collisions includes the term $-\partial_\varsigma k_e(\bm{x_l},\varsigma)/p_L(l|\lambda)f(\bm{\vec{x}'},\bm{\vec{\omega}},\varsigma)$, exactly like for true scattering events:
\begin{equation}
\begin{array}{rl}
&\partial_{\varsigma}f(\bm{\vec{x}},\bm{\vec{\omega}},\varsigma)= \displaystyle \int_{0}^{+\infty}p_{\hat{\Lambda}}(\lambda)\mathrm{d}\lambda\\
&\left(
\begin{array}{rl}
&\mathcal{H}(\lambda-\lambda_{\partial\Omega})\displaystyle \int_{0}^{\lambda_{\partial\Omega}} p_L(l|\lambda_{\partial\Omega})  \mathrm{d}l 
\left(- w_{\partial\Omega}  \frac{\partial_\varsigma k_e(\bm{x_l},\varsigma)}{p_L(l|\lambda_{\partial\Omega})}
+ w^\varsigma_{\partial\Omega}\right)\\
+&\mathcal{H}(\lambda_{\partial\Omega}-\lambda)\displaystyle \int_{0}^{\lambda} p_L(l|\lambda) \mathrm{d}l
\left(
\begin{array}{rl}
&P_A(\bm{\vec{x}'},\varsigma) \left(- w_A  \frac{\partial_\varsigma k_e(\bm{x_l},\varsigma)}{p_L(l|\lambda)}  + w^\varsigma_A\right) \\ 
+ &P_S(\bm{\vec{x}'},\varsigma)\displaystyle \int_{4\pi} p_S(-\bm{\vec{\omega}'}|-\bm{\vec{\omega}})\mathrm{d}\omega'\\
&\hspace{2cm}\begin{array}{rl}
&\left(
\begin{array}{rl}
-&\frac{\partial_\varsigma k_e(\bm{x_l},\varsigma)}{p_L(l|\lambda)}f(\bm{\vec{x}'},\bm{\vec{\omega}'},\varsigma)\\
+&\frac{\partial_\varsigma k_s(\bm{\vec{x}'},\varsigma)}{k_s(\bm{\vec{x}'},\varsigma)}f(\bm{\vec{x}'},\bm{\vec{\omega}'},\varsigma)\\
+& \partial_\varsigma f(\bm{\vec{x}'},\bm{\vec{\omega}'},\varsigma)
\end{array}
\right)
\end{array}\\
+& P_N(\bm{\vec{x}'},\varsigma) 
\left(
\begin{array}{rl}
-&\frac{\partial_\varsigma k_e(\bm{x_l},\varsigma)}{p_L(l|\lambda)}f(\bm{\vec{x}'},\bm{\vec{\omega}},\varsigma)\\
+&\partial_\varsigma f(\bm{\vec{x}'},\bm{\vec{\omega}},\varsigma)
\end{array}
\right)
\end{array}
\right)
\end{array}
\right)
\end{array}
\label{Form_int_sans_khat4}
\end{equation}
Equations~\ref{Form_int_avec_khat1} and \ref{Form_int_sans_khat4} have now a similar structure:  all the samples used to evaluate $f$ can also be used for the evaluation of $\partial_\varsigma f$. But in order to complete the evaluation of sensitivity, we must add one sample (of $L$) per collision. Thanks to this similar structure, we can gather them into a single vectorial writting (exactly the same way Eq.~\ref{Form_int_avec_khat2} was constructed from Eq.~\ref{Form_int_avec_khat1} and \ref{Form_int_s_avec_khat1}):
%
%\begin{adjustwidth}{-2cm}{0cm}
\begin{equation}
\begin{array}{rl}
&\left\{f(\bm{\vec{x}},\bm{\vec{\omega}},\varsigma)~;~\partial_{\varsigma}f(\bm{\vec{x}},\bm{\vec{\omega}},\varsigma)\right\} = \displaystyle \int_{0}^{+\infty}p_{\hat{\Lambda}}(\lambda)\mathrm{d}\lambda\\
&\left(
\begin{array}{rl}
&\mathcal{H}(\lambda-\lambda_{\partial\Omega})\displaystyle \int_{0}^{\lambda_{\partial\Omega}} p_L(l|\lambda_{\partial\Omega})  \mathrm{d}l 
\left\{ w_{\partial\Omega}~;~
- w_{\partial\Omega} \frac{\partial_\varsigma k_e(\bm{x_l},\varsigma)}{p_L(l|\lambda_{\partial\Omega})}
+ w^\varsigma_{\partial\Omega}
\right\}\\
+&\mathcal{H}(\lambda_{\partial\Omega}-\lambda)\displaystyle \int_{0}^{\lambda} p_L(l|\lambda) \mathrm{d}l
\left(
\begin{array}{rl}
&P_A(\bm{\vec{x}'},\varsigma) \left\{ w_A~;~- w_A \frac{\partial_\varsigma k_e(\bm{x_l},\varsigma)}{p_L(l|\lambda)}  + w^\varsigma_A \right\}\\ 
+ &P_S(\bm{\vec{x}'},\varsigma)\displaystyle \int_{4\pi} p_S(-\bm{\vec{\omega}'}|-\bm{\vec{\omega}})\mathrm{d}\omega'\\
&\hspace{2cm}\begin{array}{rl}
&\left\{
f(\bm{\vec{x}'},\bm{\vec{\omega}'},\varsigma)~;~
\left(
\begin{array}{rl}
-&\frac{\partial_\varsigma k_e(\bm{x_l},\varsigma)}{p_L(l|\lambda)}f(\bm{\vec{x}'},\bm{\vec{\omega}'},\varsigma) \\
+&\frac{\partial_\varsigma k_s(\bm{\vec{x}'},\varsigma)}{k_s(\bm{\vec{x}'},\varsigma)}f(\bm{\vec{x}'},\bm{\vec{\omega}'},\varsigma)\\
+& \partial_\varsigma f(\bm{\vec{x}'},\bm{\vec{\omega}'},\varsigma)
\end{array}
\right)
\right\} 
\end{array}\\
+& P_N(\bm{\vec{x}'},\varsigma)\left\{ f(\bm{\vec{x}'},\bm{\vec{\omega}},\varsigma) ~;~
\left(
\begin{array}{rl}
-&\frac{\partial_\varsigma k_e(\bm{x_l},\varsigma)}{p_L(l|\lambda)}f(\bm{\vec{x}'},\bm{\vec{\omega}},\varsigma)\\
+&\partial_\varsigma f(\bm{\vec{x}'},\bm{\vec{\omega}},\varsigma)
\end{array}
\right)
\right\}
\end{array}
\right)
\end{array}
\right)
\end{array}
\label{Form_int_sans_khat5}
\end{equation}
%\end{adjustwidth}
%
The algorithmic-reading of \ref{Form_int_sans_khat5} leads to Alg.~\ref{alg:AlgoDf2} that is an alternative to Alg.~\ref{alg:AlgoDf1} for evaluating simultaneously $f$ and $\partial_{\varsigma}f$. As explained in the algorithmic reading of Eq.\ref{Form_int_avec_khat2}, the recursive nature of this algorithm comes from the fact that the final brackets in the scattering and null-collision terms contain $f$ and $\partial_\varsigma f$ at the same location in the same direction. The fact that their sensitivity part includes a summation is translated into an algorithm incrementing the Monte Carlo weight as explained in \ref{appendix:weight}.
It is interesting to note that the integral $\int p_{L}(l)\mathrm{d}l~\partial_{\varsigma}k_e(\bm{x_l},\varsigma)/p_{L}(l)$ can be more or less difficult to evaluate depending on the profile of $\partial_\varsigma k_e$. But this can be easily handled using importance sampling based on the $\hat{k}$-adjustment grid, as explained in  \ref{appendix:additional_sampling}.

\section{Simulations using the alternative approach}
\label{sec:simulation-examples}

\subsection{Transmissivity of a purely absorbing column}
\label{subsec:transmissivity-evaluation}

Applying the alternative approach of Sec.~\ref{sec:alternative-approach} to the evaluation of column-transmissivities leads to the content of Fig.~\ref{hom_algo3} and Table~\ref{table3}. For the \emph{homogeneous-slab} configuration, Fig.~\ref{homogeneous_slab} shows that not only the pathological behavior of Sec.~\ref{sec:convergence-difficulties} is removed, but the sensitivity is estimated with a statistical uncertainty that is perfect for a simultaneous evaluation: its dependence on the parameters of the problem is identical to that of the main quantity. As above, in this very simple case, this uncertainly can be expressed analytically (see \ref{appendix_anal_var_algo_3}) and indeed $$\frac{\sigma_{\partial_{\varsigma}T}}{\partial_{\varsigma}T} = \frac{\sigma_T}{T} = \frac{\sqrt{1-e^{-k_a L}}}{\sqrt{N}}$$
For the \emph{heterogeneous-slab} configuration, the uncertainty cannot be predicted theoretically, but the conclusions of Fig.~\ref{Table} are identical to those of Fig.~\ref{homogeneous_slab}: in terms of relative accuracy, the convergence rate is equal to that of the algorithm evaluating the main quantity. It is therefore strictly independent of the adjustment of $\hat{k}$ to $k$. The use of an acceleration grid does only what we expect: it reduces the number of null-collisions but does not impact the variance anymore.

\subsection{Full radiative transfer in a 3D configuration}
\label{subsec:3D-radiative-transfer}

\begin{figure}
\begin{center}
\includegraphics[page=1,clip,scale=1]{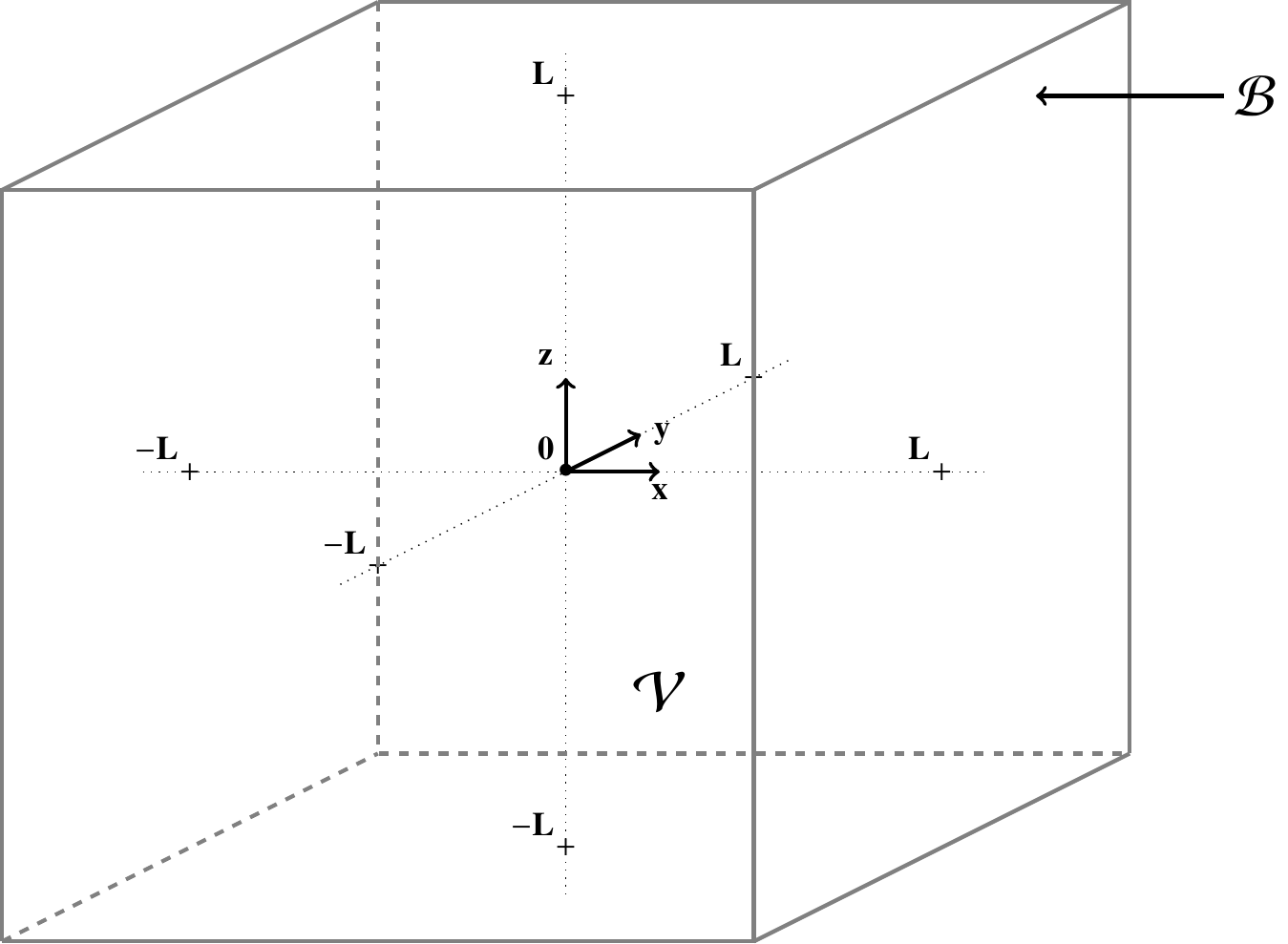}
\caption{
The \emph{heterogeneous-cube} configuration: the cube is of side $2L$, centred on $\vec{O}$, with $0K$ black faces perpendicular to the three axes of a cartesian coordinate system $(\vec{O},x,y,z)$. The inside-temperature field is such that $f^{eq}$ varies from $f^{eq}=f^{eq}_{max}$ (at the center of the face at $x=-L$) to $f^{eq}=0$ (at $x=L$ and $(y=\pm L,z=\pm L)$): $f^{eq}(x,y,z)= \eta(x,y,z) f^{eq}_{max}$ with $\eta(x,y,z) = \frac{L-x}{2L}\left(1-\sqrt{\frac{y^2+z^2}{2L^2}}\right)$. The fields of absorption and scattering coefficients follow the same spatial dependence: $k_a(x,y,z)= \eta(x,y,z) k_{a,max}$ and $k_a(x,y,z)= \eta(x,y,z) k_{s,max}$. The single-scattering phase function is that of Henyey-Greenstein with a uniform value of the asymmetry parameter g. The total extinction coefficient $\hat{k}_n$ is chosen constant in pieces on a $n\times n\times n$-regular grid, exactly equal to the maximum $k_e$-value inside each mesh. In particular, $\hat{k}_1$ is uniform on the whole cube and equal to $k_{a,max}+k_{s,max}$.
}
\label{cube}
\end{center}
\end{figure}

In \citep{galtier2013}, a cubic benchmark configuration was used to test null-collision algorithms when dealing with three-dimension highly-heterogeneous fields for all ranges of optical thickness and single-scattering albedo. We here make use of the same configuration, named \emph{heterogeneous-cube} hereafter, in order to test our alternative approach with 3D radiation (see Fig.~\ref{cube}):
\begin{itemize}
\item radiation is monochromatic;
\item the cube is of side $2L$, with $0K$ black faces;
\item the inside-temperature field is such that $f^{eq}$ varies from $f^{eq}=f^{eq}_{max}$ (at the center of the face at $x=-L$) to $f^{eq}=0$ ({at $x=L$ and $(y=\pm L,z=\pm L)$}) and mimics the shape of flame: $f^{eq}(x,y,z)=\eta(x,y,z)f^{eq}_{max}$ (see the $\eta$ profile in Fig.~\ref{cube});
\item the fields of absorption and scattering coefficients follows the same spatial dependence: $k_a(x,y,z)= \eta(x,y,z) k_{a,max}$ and $k_a(x,y,z)= \eta(x,y,z) k_{s,max}$;
\item The single-scattering phase function is that of Henyey-Greenstein with a uniform value of the asymmetry parameter g;
\item $\hat{k}$ is adjusted to $k$ using a regular cubic-grid ($\hat{k}$ uniform within each mesh): the only parameter for $\hat{k}$ is therefore the number of mesh per direction.
\end{itemize}
The evaluated quantity $A(x,y,z)$ is the stationary net-power density and the free physical parameters are $k_{a,max}L$, $k_{s,max}L$ and $g$. In Table~\ref{Table_cube} we reproduce the computations of Table~1 in \citep{galtier2013}, i.e. testing wide ranges of optical thicknesses but fixing $g=0$ (isotropic scattering). In the same table we also provide two sensitivities, $\partial_{k_{a,max}}$ and $\partial_{k_{s,max}}$, that we evaluated simultaneously with $A$. As in \citep{galtier2013}, although they are not displayed, we checked that simulation results with non-isotropic scattering lead to the exact same conclusions.

These conclusions are very similar to those reached on the slab-transmissivity example: Table~\ref{Table1_cube} highlights the same features as Fig.~\ref{Table}, the standard deviation of Alg.~\ref{alg:AlgoDf2} being independent of $\hat{k}$, whereas it increases when $\hat{k}$ gets close to $k_e$ for Alg.~\ref{alg:AlgoDf1}.

\section{Conclusion}
\label{sec:conclusion}

The simulation examples of the preceding section indicate that the solution proposed in Sec.~\ref{sec:alternative-approach} is practical. Sensitivities can be accurately evaluated even when using null-collision algorithms. We still face convergence difficulties for highly-scattering media, but this is an open question identified in all linear-transport physics, independently of the use of null-collisions.

The way we bypassed the convergence difficulties specific to null-collisions is in rupture with the principle of evaluating sensitivities $\partial_\varsigma A$ simultaneously with the main quantity $A$: in all previous works, the two evaluations were truly simultaneous in the sense that the very same samples were used in the $A$ and $\partial_\varsigma A$ Monte Carlo algorithms. Here, all the samples used to evaluate $A$ are also used for the evaluation of $\partial_\varsigma A$, but in order to complete the evaluation of $\partial_\varsigma A$, some additional random variables must be sampled. The sensitivity evaluation has therefore a specific computation-cost. In all our test-cases, this cost was very small compared to the total computation cost, and considering the wide range of tested optical depths we may state that this additional cost  has no practical significance for radiative transfer applications. Other linear-transport physicas wold have to be investigated specifically.

In the present text, we only gave very little indications about absolute computer-requirements. We provided some computation-time examples in Fig.~\ref{Table_cube}, without the use of any acceleration grid. Most of our analysis was made with another measure: the average number of random generation per path, which should be proportional to the computation time, but only once all optimisations will be implemented. As illustrated in \citep{kutz2017,novak2014,novak2018,raab2008}, such data-access considerations in relation with null-collision algorithms (i.e the time associated to a close adjustment of $\hat{k}$ to $k$) is an ongoing computer-science research. The question of implementing the corresponding tools (developped by the computer-graphics community) together with the present sensitivity evaluation algorithm will be discussed in a separate paper.

%
%%%%%%%%%%%%%%%%%%%%%%%%%%%%%%%%%%%%%%%%%%%%%%%APPENDIX%%%%%%%%%%%%%%%%%%%%%%%%%%%%%%%%%%%%%%%%%%%%%%%
\appendix
\section{Analytical variances for \emph{homogeneous-slab}}
\label{appendix:homogeneous-slab}
$\Omega$ is a slab of length $L$ with $\bm{\vec{e_x}}$ the incoming normal at location $\bm{\vec{y}}=\bm{0}$ and $f_{\partial\Omega}(L,-\bm{\vec{e_x}},.)=f_{inc}$ as boundary condition (see figure \ref{fig1}). The equilibrium distribution $f^{eq}$ is null (no emssion) and the absorption and scattering coefficients are $k_a\equiv \varsigma$ and $k_s\equiv0$. The transmissivity is $T=f(\bm{0},\bm{\vec{e_x}},.,\varsigma)/f_{inc}$.
\subsection{For the standard approachs}
\label{appendix_anal_var_algo_2}
Using Eq.\eqref{Form_int_avec_khat2},
\begin{equation}
\begin{split}
\partial_{\varsigma}T(L,\varsigma)=& \displaystyle \int_{0}^{\infty} \hat{k} e^{-\hat{k}\lambda} \, \mathrm{d}\lambda
\left\{
\begin{array}{rl}
 &\mathcal{H}(\lambda-L) ~0 \\
 +&\mathcal{H}(L-\lambda) 
  \left\{
\begin{array}{rl}
&P_A(\varsigma)~0\\
+&P_N(\varsigma)\Big( \frac{-1}{k_n(\varsigma)} T(L-\lambda,\varsigma) + \partial_{\varsigma}T(L-\lambda,\varsigma)\Big)\Big)
 \end{array}
\right\}
\end{array}
\right\}
\end{split}
\label{DTran_hom_meth1}
\end{equation}
~\\
To get the analytical variance for the standard approach (Alg.~\ref{alg:AlgoDf1}) in \emph{homogeneous-slab}, we started by designing a countable categorisation of paths: $\mathcal{C}_n $ ($ n\in\mathbb{N}^* $) is the set of paths that crossed the column and met $ n $ null-collisions and $ \mathcal{C} $ all other paths (path without null-collisions).
The probability $ p_n $ that a path belongs to the $\mathcal{C}_n $ category is
\begin{center}
$p_n=\frac{\hat{k}^n}{n!}L^n \mathrm{e}^{-\hat{k}L}\Big(1-\frac{k_a(\varsigma)}{\hat{k}}\Big)$
\end{center}
The Monte Carlo weight $ w_n $ obtained at the end of a path belonging to the $\mathcal{C}_n $ is
\begin{center}
$w_n=\frac{-n}{k_n}$
\end{center}
The weight of a path belonging to $ \mathcal{C} $ is zero. Therefore, the variance of the  Alg.~\ref{alg:AlgoDf1} is
\begin{equation}
\begin{array}{rl}
Var_{Alg.\ref{alg:AlgoDf1}}(\partial_{\varsigma}T(L,\varsigma))&= \sum_{n = 1}^{\infty} p_n w_n^2 - \Big(\sum_{n = 1}^{\infty} p_n w_n\Big)^2\\
&=L^2\mathrm{e}^{-k_a L} \Big(\frac{k_n +1/L}{k_n}\Big)- L^2 \mathrm{e}^{-2 k_a L}
\end{array}
\end{equation}
\subsection{For the alternative approach}
\label{appendix_anal_var_algo_3}
Using Eq.\eqref{Form_int_sans_khat5},
\begin{equation}
\begin{split}
\partial_{\varsigma}T(L,\varsigma)=& \displaystyle \int_{0}^{\infty} \hat{k} e^{-\hat{k}\lambda} \, \mathrm{d}\lambda
\left\{
\begin{array}{rl}
 &\mathcal{H}(\lambda-L) ~(-L) \\
 +&\mathcal{H}(L-\lambda) 
  \left\{
\begin{array}{rl}
&P_A(\varsigma)~0\\
+&P_N(\varsigma)~\partial_{\varsigma}T(L-\lambda,\varsigma)
 \end{array}
\right\}
\end{array}
\right\}
\end{split}
\label{DTran_hom_meth1}
\end{equation}
The variance of a Bernoulli law of $p$ parameter is $p(1-p)$, and therefore the variance of Alg.~\ref{alg:AlgoDf2} is 
\begin{equation}
Var_{Alg.\ref{alg:AlgoDf2}}(\partial_{\varsigma}T(L,\varsigma))=L^2\mathrm{e}^{-k_a L}\left(1-\mathrm{e}^{-k_a L}\right)
\end{equation}
Similarly for Alg.~\ref{alg:Algof},
\begin{equation}
Var_{Alg.\ref{alg:Algof}}(\partial_{\varsigma}T(L,\varsigma))=\mathrm{e}^{-k_a L}\left(1-\mathrm{e}^{-k_a L}\right)
\end{equation}
~\\
\section{Incrementation of the Monte Carlo weights}
\label{appendix:weight}
The standard approach of Alg.~\ref{alg:AlgoDf1} is designed from the algorithmic-reading of Eq.~\ref{Form_int_avec_khat2} and the recursive nature of this algorithm comes from the fact that the final brackets in the scattering and null-collision terms contain $f$ and $\partial_\varsigma f$ at the same location in the same direction. As the integral formulation of $f$ and $\partial_\varsigma f$ have identical structures, we can just use one unique sampled-path to evaluate simultaneously $f$ and $\partial_\varsigma f$.
For a better understanding of the meaning of this unique sampled-path, we express here the Monte Carlo weight for one path example: the path displayed in Fig.~\ref{fig_Algof}, i .e. one null-collision at location $\vec{\bm{x_1}}$, two scatterings at location $\vec{\bm{x_2}}$ and $\vec{\bm{x_3}}$, and one boundary-collision at location $\vec{\bm{y_4}}$. Thanks to the algorithmic-reading of Eq.~\ref{Form_int_avec_khat2}, after one null-collision at location $\vec{\bm{x_1}}$. The notations are simplified by writing $"f"$ instead of $w_{"f"}$. We have
\begin{equation}
\partial_\varsigma f(\vec{\bm{x_0}},\vec{\bm{\omega_0}},\varsigma)=\partial_\varsigma f(\vec{\bm{x_1}},\vec{\bm{\omega_0}},\varsigma) + \frac{\partial_\varsigma k_n(\vec{\bm{x_1}},\varsigma)}{k_n(\vec{\bm{x_1}},\varsigma)}f(\vec{\bm{x_1}},\vec{\bm{\omega_0}},\varsigma)
\end{equation}
At this stage, $\partial_\varsigma f$ and $f$ need to be evaluated at the same location ($\vec{\bm{x_1}}$) in the same direction ($\vec{\bm{\omega_0}}$). According to Eq.~\ref{Form_int_avec_khat2}, it's possible to use the same path to evaluate $\partial_\varsigma f$ and $f$. Therefore, after one scattering-collision at location $\vec{\bm{x_2}}$ in direction $\vec{\bm{\omega_2}}$, we have $\partial_\varsigma f(\vec{\bm{x_1}},\vec{\bm{\omega_0}},\varsigma)= \partial_\varsigma f(\vec{\bm{x_2}},\vec{\bm{\omega_2}},\varsigma)+\frac{\partial_\varsigma k_s(\vec{\bm{x_2}},\varsigma)}{k_s(\vec{\bm{x_2}},\varsigma)}f(\vec{\bm{x_2}},\vec{\bm{\omega_2}},\varsigma)$ and $f(\vec{\bm{x_1}},\vec{\bm{\omega_0}},\varsigma)= f(\vec{\bm{x_2}},\vec{\bm{\omega_2}},\varsigma)$,
\begin{equation}
\begin{array}{rl}
\partial_\varsigma f(\vec{\bm{x_0}},\vec{\bm{\omega_0}},\varsigma)=&\partial_\varsigma f(\vec{\bm{x_2}},\vec{\bm{\omega_2}},\varsigma) +\frac{\partial_\varsigma k_s(\vec{\bm{x_2}},\varsigma)}{k_s(\vec{\bm{x_2}},\varsigma)}f(\vec{\bm{x_2}},\vec{\bm{\omega_2}},\varsigma)+ \frac{\partial_\varsigma k_n(\vec{\bm{x_1}},\varsigma)}{k_n(\vec{\bm{x_1}},\varsigma)}f(\vec{\bm{x_2}},\vec{\bm{\omega_2}},\varsigma)\\
=&\partial_\varsigma f(\vec{\bm{x_2}},\vec{\bm{\omega_2}},\varsigma) +\left(\frac{\partial_\varsigma k_s(\vec{\bm{x_2}},\varsigma)}{k_s(\vec{\bm{x_2}},\varsigma)}+ \frac{\partial_\varsigma k_n(\vec{\bm{x_1}},\varsigma)}{k_n(\vec{\bm{x_1}},\varsigma)}\right)f(\vec{\bm{x_2}},\vec{\bm{\omega_2}},\varsigma)
\end{array}
\end{equation}
Similarly, after one scattering at location $\vec{\bm{x_3}}$ in direction $\vec{\bm{\omega_3}}$,
\begin{equation}
\begin{array}{rl}
\partial_\varsigma f(\vec{\bm{x_0}},\vec{\bm{\omega_0}},\varsigma)=&\partial_\varsigma f(\vec{\bm{x_3}},\vec{\bm{\omega_3}},\varsigma) + \frac{\partial_\varsigma k_s(\vec{\bm{x_3}},\varsigma)}{k_s(\vec{\bm{x_3}},\varsigma)}f(\vec{\bm{x_3}},\vec{\bm{\omega_3}},\varsigma) +\left(\frac{\partial_\varsigma k_s(\vec{\bm{x_2}},\varsigma)}{k_s(\vec{\bm{x_2}},\varsigma)}+ \frac{\partial_\varsigma k_n(\vec{\bm{x_1}},\varsigma)}{k_n(\vec{\bm{x_1}},\varsigma)}\right)f(\vec{\bm{x_3}},\vec{\bm{\omega_3}},\varsigma)\\
=&\partial_\varsigma f(\vec{\bm{x_3}},\vec{\bm{\omega_3}},\varsigma) + \left(\frac{\partial_\varsigma k_s(\vec{\bm{x_3}},\varsigma)}{k_s(\vec{\bm{x_3}},\varsigma)} +\frac{\partial_\varsigma k_s(\vec{\bm{x_2}},\varsigma)}{k_s(\vec{\bm{x_2}},\varsigma)}+ \frac{\partial_\varsigma k_n(\vec{\bm{x_1}},\varsigma)}{k_n(\vec{\bm{x_1}},\varsigma)}\right)f(\vec{\bm{x_3}},\vec{\bm{\omega_3}},\varsigma)
\end{array}
\end{equation}
and after the boundary-collision at location $\vec{\bm{y_4}}$ in direction $\vec{\bm{\omega_3}}$,
\begin{equation}
\begin{array}{rl}
\partial_\varsigma f(\vec{\bm{x_0}},\vec{\bm{\omega_0}},\varsigma)=&w^\varsigma_{\partial\Omega} +  \left( \frac{\partial_\varsigma k_s(\vec{\bm{x_3}},\varsigma)}{k_s(\vec{\bm{x_3}},\varsigma)} +\frac{\partial_\varsigma k_s(\vec{\bm{x_2}},\varsigma)}{k_s(\vec{\bm{x_2}},\varsigma)}+ \frac{\partial_\varsigma k_n(\vec{\bm{x_1}},\varsigma)}{k_n(\vec{\bm{x_1}},\varsigma)}\right) w_{\partial\Omega}
\end{array}
\end{equation}
where $w^\varsigma_{\partial\Omega}=0$ and $w_{\partial\Omega}=f_{\partial\Omega}(\vec{\bm{y_4}},\vec{\bm{\omega_3}},\varsigma)$.
The summation of $f$ and $\partial_\varsigma f$ is therefore translated into an algorithm incrementing the Monte Carlo weight. This is how the Monte Carlo weight is incremented in the fully recursive algorithm of Alg.~\ref{alg:AlgoDf2}.
\section{The additional sampling}
\label{appendix:additional_sampling}
To evaluate $\partial_{\varsigma}T$ with the alternative approach of Alg.~\ref{alg:AlgoDf2}, we must evaluate $\displaystyle \int_{0}^{\lambda}\partial_{\varsigma}k_e(\vec{\bm{x_l}},\varsigma)\mathrm{d}l$. We make use of one additional sampling:
\begin{equation}
\displaystyle \int_{0}^{\lambda}\partial_{\varsigma}k_e(\vec{\bm{x_l}},\varsigma)\mathrm{d}l=\displaystyle \int_{0}^{\lambda}p_{L}(l)\mathrm{d}l~\partial_{\varsigma}k_e(\vec{\bm{x_l}},\varsigma)/p_{L}(l)
\end{equation}
However, depending  on the the profile of $\partial_{\varsigma}k_e$, this integral can be source of variance if the spatial variations of $p_{L}(l)$ do not sufficiently match those $\partial_{\varsigma}k_e(\vec{\bm{x_l}}$. We did not yet adress this importance-sampling issue and bypassed it using stratified sampling: we increased the number of "collisions" by including the grid-intersections to write
\begin{equation}
\begin{array}{rl}
\displaystyle \int_{0}^{\lambda}\partial_{\varsigma}k_e(\vec{\bm{x_l}},\varsigma)\mathrm{d}l=&\displaystyle \int_{0}^{\beta_1}p_{L}(l_1)\mathrm{d}l_1 ~\partial_{\varsigma}k_e(\vec{\bm{\zeta_{1}}},\varsigma)/p_{L}(l_1)\\
+&\displaystyle \int_{0}^{\beta_2}p_{L}(l_2)\mathrm{d}l_2 ~\partial_{\varsigma}k_e(\vec{\bm{\zeta_{2}}},\varsigma)/p_{L}(l_2)\\
+&...\\
+&\displaystyle \int_{0}^{\beta_n}p_{L}(l_n)\mathrm{d}l_n ~\partial_{\varsigma}k_e(\vec{\bm{\zeta_{n}}},\varsigma)/p_{L}(l_n)
\end{array}
\label{eq_coll_grid}
\end{equation}   
where $n$ is the number of grid-intersections along the path $[\vec{\bm{x_0}},\vec{\bm{x_\lambda}}]$. For all $i\in[1,n]$, $\vec{\bm{z_{i}}}$ is the location of the $ith$ "collision", $\beta_i$ is the length between two successive "collisions" and $\vec{\bm{\zeta_i}}=\vec{\bm{z_i}}-l_i\vec{\bm{\omega}}$ (see Fig.~\ref{fig_col_grid}).

This stratified sampling is only meaningful if we assume that the grid is sufficiently refined to capture the spatial variation of $\partial_{\varsigma}k_e(\vec{\bm{x_l}})$. But initially, the grid was only designed to reduce the amount of null-collisions. Therefore, another grid-refinement criterium should be introduced if this strategy is retained. But again, stratified sampling is not the only possible solution and the grid may be used differently: make pre-computations in each mesh and sample $L$ only once along the whole path using importance sampling, which would suppress the increase of computation costs when refining the grid in Fig.~\ref{Table} ($R$ increasing in Fig.~\ref{table3}).
%%%%%%%%%%%%%%%%%%%%%%%%
%%%%%%%%%%%%%%%%%%%%%%%%
%%%%%%%%%%%%%%%%%%%%%%%%
%%%%%%%%%%%%%%%%%%%%%%%%
%%%%%%%%%%%%%%%%%%%%%%%%
%%%%%%%%%%%%%%%%%%%%%%%%
%%%%%%%%%%%%%%%%%%%%%%%%
%%%%%%%%%%%%%%%%%%%%%%%%
%%%%%%%%%%%%%%%%%%%%%%%%
%%%%%%%%%%%%%%%%%%%%%%%%
%%%%%%%%%%%%%%%%%%%%%%%%
%%%%%%%%%%%%%%%%%%%%%%%%
%%%%%%%%%%%%%%%%%%%%%%%%

\section{Another possible approach: $\hat{k}$ varying with $k_e$}
\label{appendix-k-chapeau-variable}
In Eq. \eqref{Form_int_avec_khat1}, we chose to use a constant upper bound of $ k_e $ noted $\hat{k}$. Nevertheless, nothing prevents us from varying $ \hat{k} $ according to the parameter $\varsigma$, while preserving the fact that $ \hat{k}(.,\varsigma) $ is independent of space in order to still enable an easy sampling of $\hat{\Lambda}$. Thus, we have complete freedom in the choice of these variations, and we will choose them to reduce the variance as much as possible.\\
In addition, we have shown that the explosion is due to the fact that Monte Carlo's weight includes a factor $ 1/k_n $. This factor appeared when applying the sensitivity evaluation technique to Eq. \eqref {Form_int_avec_khat1}, especially because of the term $ \partial_{\varsigma} \Big(\frac{k_n} {\hat{k}} \Big) $. \\
Therefore, and now that $ \hat{k} $ depends on $ \varsigma $, we may impose $ \partial_{\varsigma} \Big(\frac{k_n} {\hat{k}} \Big) = 0 $, which implies the following relation:
$$\partial_{\varsigma}\hat{k}=\frac{\partial_{\varsigma}k_e~\hat{k}}{k_e}
$$
Thus, by derivating Eq.~\eqref{Form_int_avec_khat1}, we obtain:
%
%\begin{adjustwidth}{-2cm}{0cm}
\begin{equation}
\begin{array}{rl}
\partial_{\varsigma}f(\bm{\vec{x}},\bm{\vec{\omega}},\varsigma)=&\\
\displaystyle \int_{0}^{+\infty}p_{\hat{\Lambda}}(\lambda)\mathrm{d}\lambda&\left(
\begin{array}{rl}
&\mathcal{H}(\lambda-\lambda_{\partial\Omega})\left[-w_{\partial\Omega}\displaystyle \int_0^{\lambda_{\partial\Omega}} \partial_{\varsigma}\hat{k}(\bm{\vec{x}_l},\varsigma) \mathrm{d}l  + w_{\partial\Omega}^{\varsigma}\right]\\
+&\mathcal{H}(\lambda_{\partial\Omega}-\lambda)
\left(
\begin{aligned}
&P_A(\bm{\vec{x}'},\varsigma)\left[\begin{array}{rl}
&\psi(\bm{\vec{x}},\lambda,\varsigma)w_A\\
+& \left(\frac{\partial_{\varsigma}k_a}{k_a}-\frac{\partial_{\varsigma}\hat{k}}{\hat{k}}\right)_{|\bm{\vec{x}'},\varsigma}w_A\\
+&w_A^{\varsigma}
\end {array}\right]\\ 
+&P_S(\bm{\vec{x}'},\varsigma)\displaystyle \int_{4\pi} p_S(-\bm{\vec{\omega}'}|-\bm{\vec{\omega}})\, \mathrm{d}\omega'\left[\begin{array}{rl}
&\psi(\bm{\vec{x}},\lambda,\varsigma)f(\bm{\vec{x}'},\bm{\vec{\omega}'},\varsigma)\\
+& \left(\frac{\partial_{\varsigma}k_s}{k_s}-\frac{\partial_{\varsigma}\hat{k}}{\hat{k}}\right)_{|\bm{\vec{x}'},\varsigma}f(\bm{\vec{x}'},\bm{\vec{\omega}'},\varsigma)\\
+&\partial_{\varsigma}f(\bm{\vec{x}'},\bm{\vec{\omega}'},\varsigma)
\end{array}\right] \\ 
+&P_N(\bm{\vec{x}'},\varsigma)\left[\begin{array}{rl}
&\psi(\bm{\vec{x}},\lambda,\varsigma)f(\bm{\vec{x}'},\bm{\vec{\omega}},\varsigma)\\
+&\partial_{\varsigma}f(\bm{\vec{x}'},\bm{\vec{\omega}},\varsigma)
\end{array}\right]
\end{aligned}
\right)
\end{array}
\right)
\end{array}
\label{Form_int_avec_khat_var}
\end{equation}
%\end{adjustwidth}

with $\psi(\bm{\vec{x}},\lambda,\varsigma)=\left(\frac{\partial_{\varsigma}\hat{k}(\bm{\vec{x}'},\varsigma)}{\hat{k}(\bm{\vec{x}'},\varsigma)}-\displaystyle \int_0^{\lambda} \partial_{\varsigma}\hat{k}(\bm{\vec{x}_l},\varsigma) \mathrm{d}l\right)$

We have implemented the algorithm resulting from this integral formulation and compared it with the approach proposed in the core of the present paper. Whatever the case, it was always less efficient.
%%%%%%%%%%%%%%%%%%%%%%%%%%%%%%%%%%%%%%%%%%%%%%%%%%%%%%%%%%%%%%%%%%%%%%%%%%%%%%%%%%%%%%%%
% Algos
%%%%%%%%%%%%%%%%%%%%%%%%%%%%%%%%%%%%%%%%%%%%%%%%%%%%%%%%%%%%%%%%%%%%%%%%%%%%%%%%%%%%%%%%
%
\begin{adjustwidth}{-2cm}{0cm}
\begin{algorithm}[p]
\begin{center}
\includegraphics[page=1,clip,scale=.7,viewport=60 140 550 730]{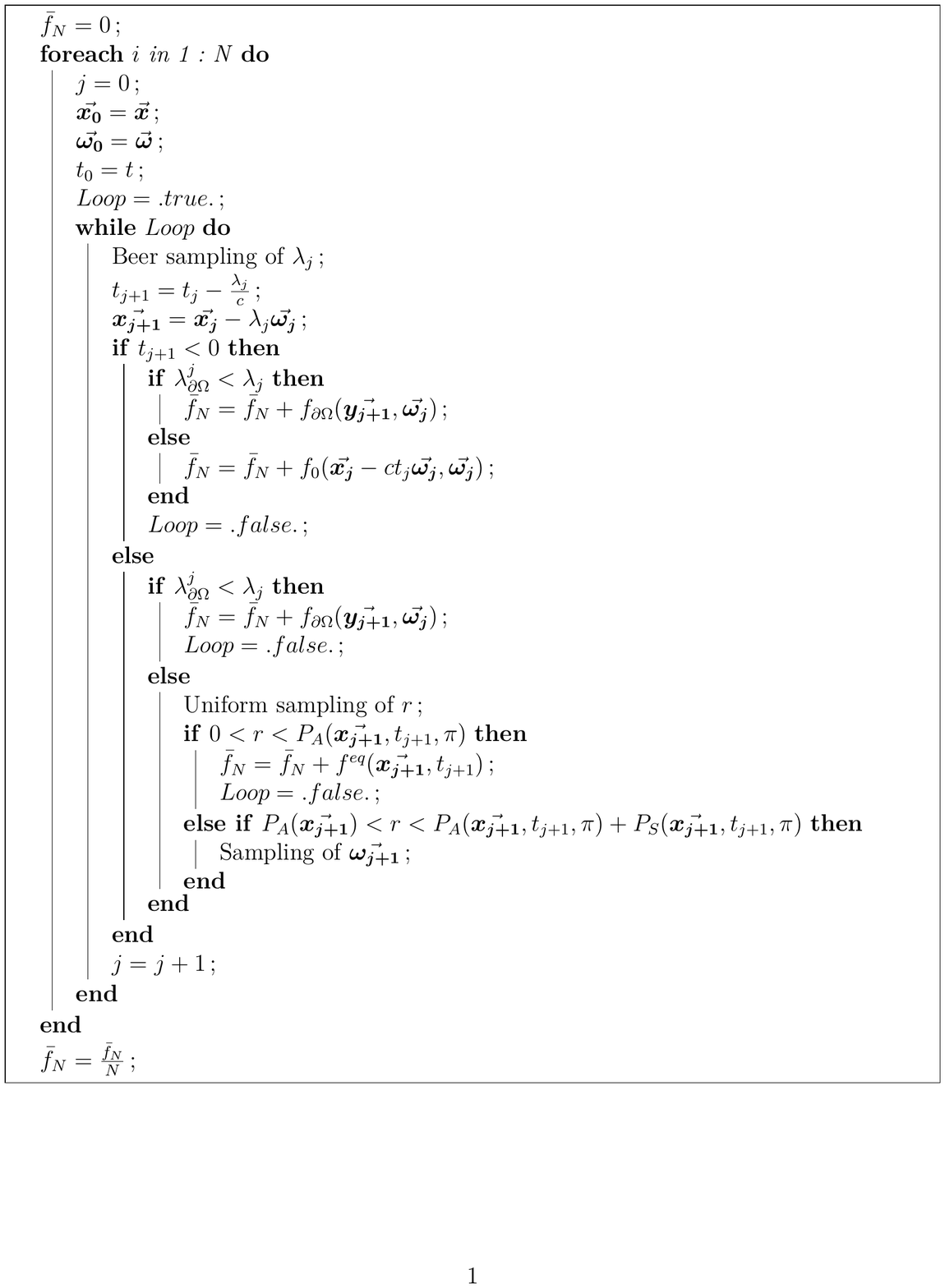}
\end{center}
\caption{
The initial backward tracking null-collision algorithm. The objective of simultaneously evaluating sensitivities will be discussed on the basis of this initial algorithm, with the idea that only small algorithmic modifications should be required to reach this goal. $N$ is the number of samples used to produce one Monte Carlo estimate of $f(\bm{\vec{x}},\bm{\vec{\omega}},t,\varsigma)$ along the algorithmic reading of Eq.\eqref{Form_int_avec_khat1}. This estimate is noted $\bar{f}_N$. $k_a$, $k_s$, $k_n$ and $\hat{k}=k_a+k_s+k_n$ are respectively the absorption coefficient, the scattering coefficient, the null-collision coefficient and the total extinction coefficient. At a given collision, $P_A(\bm{\vec{x}},t,\varsigma)=\frac{k_a(\bm{\vec{x}},t,\varsigma)}{\hat{k}}$,  $P_S(\bm{\vec{x}},t,\varsigma)=\frac{k_s(\bm{\vec{x}},t,\varsigma)}{\hat{k}}$ and $P_N(\bm{\vec{x}},t,\varsigma)=\frac{k_n(\bm{\vec{x}},t,\varsigma)}{\hat{k}}$ are respectively the absorption probability, the scattering probability and the null-collision probability. $x_0=x$ is the initial position, $\omega_0=\omega$ is the initial direction, and $t_0=t$ is the observation time. $\Omega$ is the geometrical domain, $\partial\Omega$ its boundary at which the distribution function $f_{\partial\Omega}$ is known for all locations $\bm{\vec{y}}$ and all directions $\bm{\vec{\omega}}_+$ of the incoming hemisphere $\mathbb{S}_+^2$. $f^{eq}$ is the equilibrium distribution (following the Planck function) and $f_0$ is the initial condition.  For all $j\in\mathbb{N}$, $\bm{x_{j+1}}= \bm{x_j}-\lambda_j\bm{\omega_j}$, $t_{j+1}=t_{j}-\frac{\lambda_j }{c}$, $\bm{y_{j+1}}= \bm{x_j}-\lambda_{\partial\Omega}^j\bm{\omega_j}$~ s.t. $ \lambda_{\partial\Omega}^j=\mathrm{min}\{\norme{\bm{x_j}-\bm{\vec{y}}}$;~$\bm{\vec{y}}\in\mathrm{Vect}^-(\bm{x_{j}},\bm{\omega_{j}})\cap\partial\Omega \}$ where $\mathrm{Vect}^-(\bm{\vec{x}},\bm{\vec{\omega}})=\{\bm{\vec{x}}-\lambda\bm{\vec{\omega}}$;~ $\lambda\in\mathbb{R}_+\}$.}
\label{alg:Algof}
\end{algorithm}
\FloatBarrier
\end{adjustwidth}
\begin{algorithm}[p]
\begin{center}
\includegraphics[page=1,clip,scale=.7,viewport=60 115 540 770]{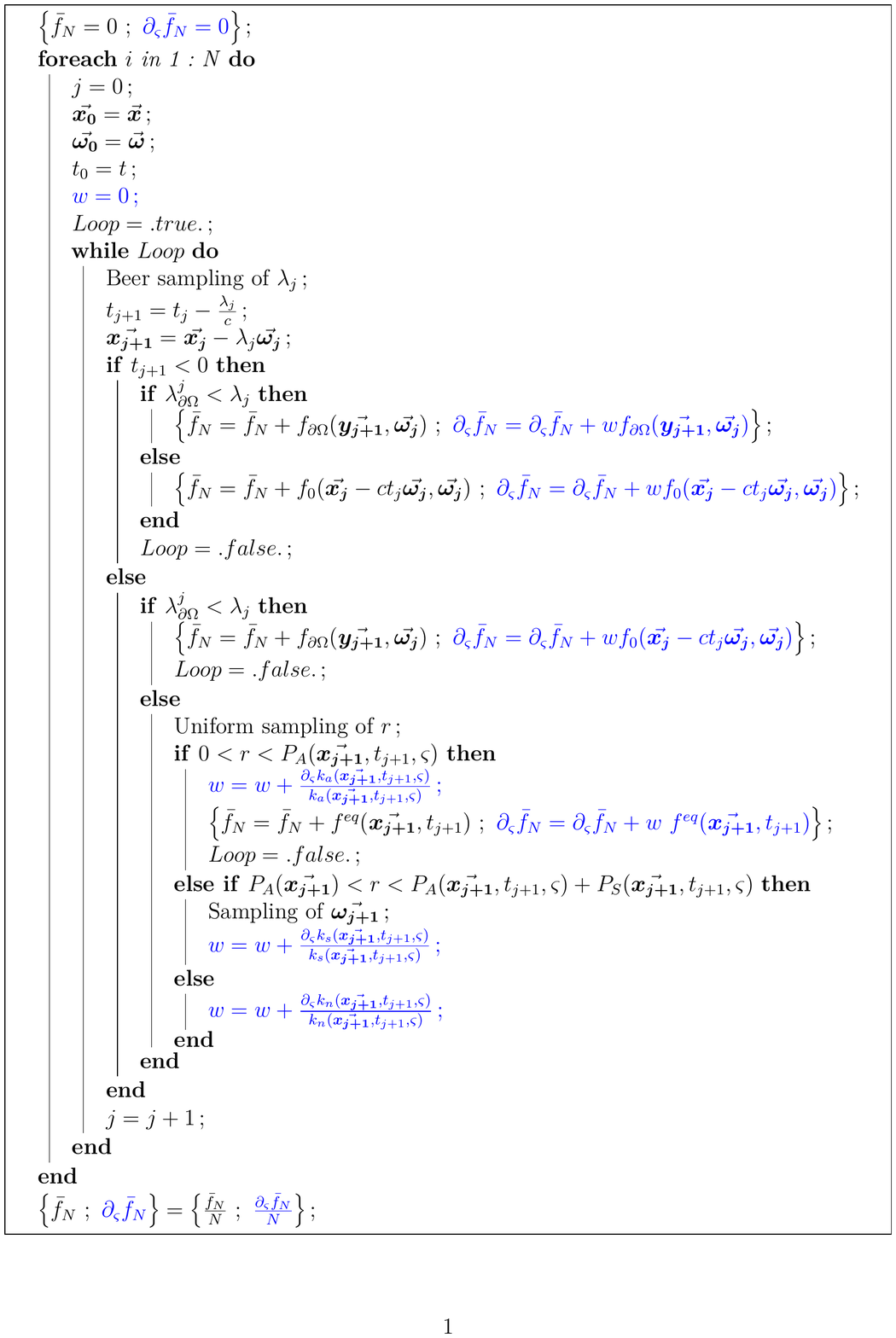}
\end{center}
\caption{
The standard approach to the simultaneous evaluation of $f$ and its sensitivity. $\partial_{\varsigma}\bar{f}_N$ is a Monte Carlo estimate of $\partial_{\varsigma}f(\bm{\vec{x}},\bm{\vec{\omega}},t,\varsigma)$ corresponding to the algorithmic reading of Eq.\eqref{Form_int_avec_khat2}. It is constructed together with $\bar{f}_N$. Notations are described inthe caption of Alg.~\ref{alg:Algof}. The incrementation of the Monte Carlo weight $w$ is detailled in Appendix~\ref{appendix:weight}}.
\label{alg:AlgoDf1}
\end{algorithm}
\FloatBarrier
\begin{algorithm}[p]
\begin{center}
\includegraphics[page=1,clip,scale=.7,viewport=30 35 590 790]{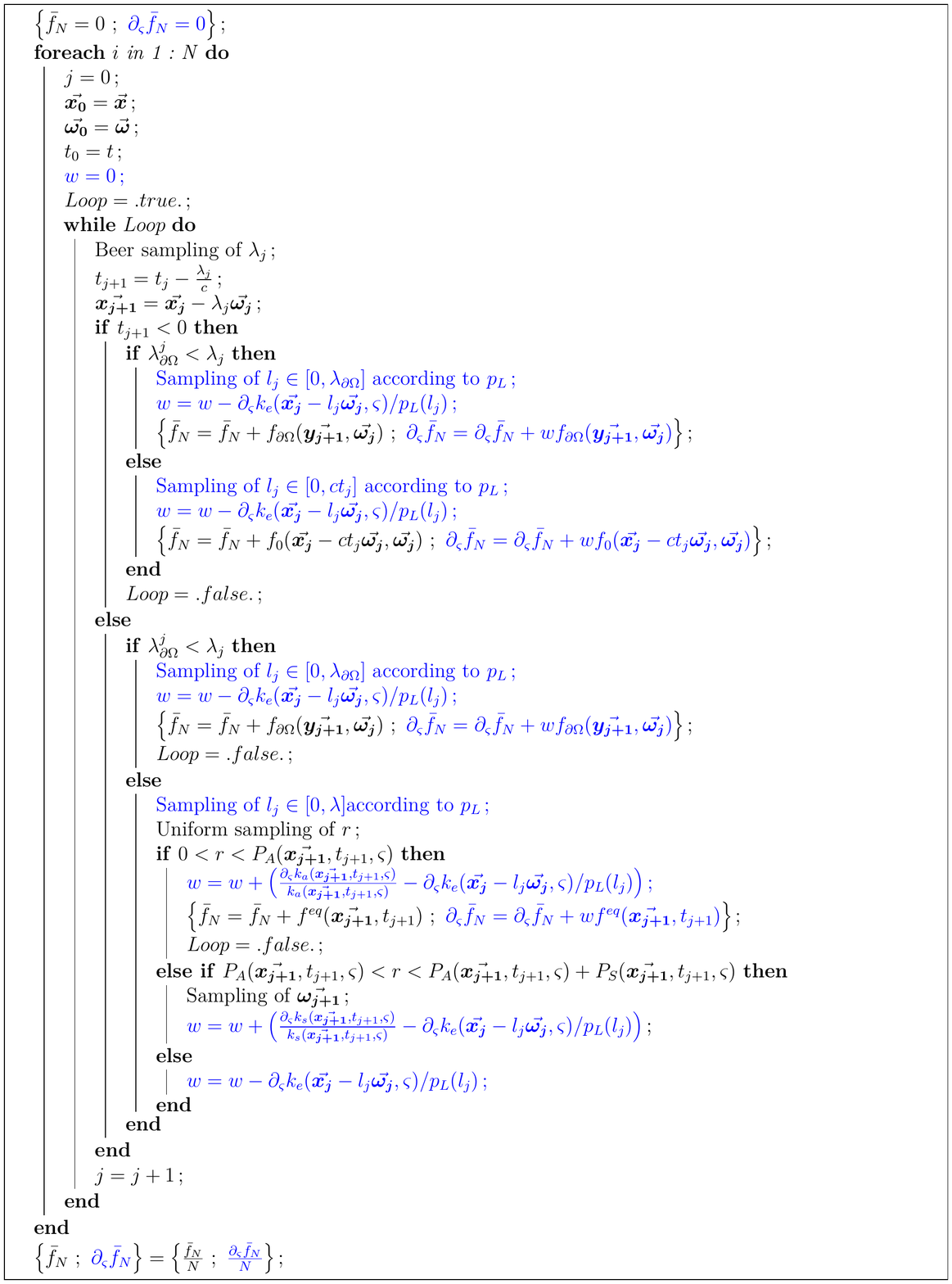}
\end{center}
\caption{
The alternative approach to the simultaneous evaluation of $f$ and its sensitivity. It corresponds to the algorithmic reading of Eq.\eqref{Form_int_sans_khat5}. Notations are described in the captions of Alg.~\ref{alg:Algof} and Alg.~\ref{alg:AlgoDf1}. The incrementation of the Monte Carlo weight $w$ is detailled in Appendix~\ref{appendix:weight}}.
\label{alg:AlgoDf2}
\end{algorithm}
\FloatBarrier
%
%%%%%%%%%%%%%%%%%%%%%%%%%%%%%%%%%%%%%%%%%%%%%%%%%%%%%%%%%%%%%%%%%%%%%%%%%%%%%%%%%%%%%%%%

%%%%%%%%%%%%%%%%%%%%%%%%%%%%%%%%%%%%%%%%%%%%%%%%%%%%%%%%%%%%%%%%%%%%%%%%%%%%%%%%%%%%%%%%
% Figure
%%%%%%%%%%%%%%%%%%%%%%%%%%%%%%%%%%%%%%%%%%%%%%%%%%%%%%%%%%%%%%%%%%%%%%%%%%%%%%%%%%%%%%%%
%
\FloatBarrier
%%%%%%%%%%%%%
%  figure slab homogène
%%%%%%%%%%%%%
\FloatBarrier
\FloatBarrier

%%%%%%%%%%%%%%%%%%%%%%%%%%%%%%%%%%%%%%%%%%%%%%%%%%%%%%%%%%%%%%%%%%%%%%%%%%%%%%%%%%%%%%%%%%%%%%
% Tables
%%%%%%%%%%%%%%%%%%%%%%%%%%%%%%%%%%%%%%%%%%%%%%%%%%%%%%%%%%%%%%%%%%%%%%%%%%%%%%%%%%%%%%%%%%%%%%

\begin{figure}[h]
  \begin{center}
    \subfloat[
Evaluation of transmissivity $T$ with Alg.~\ref{alg:Algof}.]{
\begin{tabular}{|L{1.5cm} || C{2.25cm} | C{2.25cm} | C{2.25cm} | C{2.25cm} | }
\cline{2-5}
\multicolumn{1}{c|}{}  & \multicolumn{4}{c |}{\Large{\textbf{Transmissivity}}}\\
\cline{1-5} 
$\hat{k}$  & $T$ & $\sigma$ &  $\bar{C}$ & $\bar{R}$ \\
\hline 
\hline
\cline{1-5} 
$\hat{k}_{20\%}$   & 0.327    &5.179E-03    &  116.65    &    235.2  \\
\cline{1-5} 
$\hat{k}_1$    & 0.327    & 5.358E-03    &   96.64    &    194.6  \\
\cline{1-5} 
$\hat{k}_5$    & 0.327    & 5.264E-03    &   20.06    &    38.7  \\
\cline{1-5} 
$\hat{k}_{10}$    & 0.332    & 4.984E-03    &   9.86    &    20.5 \\
\cline{1-5} 
$\hat{k}_{100}$     & 0.335    & 5.285E-03    &    1.45    &  3.4   \\
\cline{1-5} 
$\hat{k}_{1000}$      & 0.322    & 4.815E-03    &    0.72    &    1.8 \\
\hline
\end{tabular}%      
\label{table1}
                         }\\
    \subfloat[Evaluation of sensitivity $\partial_{\varsigma}T$ with Alg.~\ref{alg:AlgoDf1}.]{
\begin{tabular}{|L{1.5cm} || C{2.25cm} | C{2.25cm} | C{2.25cm} | C{2.25cm} | }
\cline{2-5}
\multicolumn{1}{c|}{}  & \multicolumn{4}{c |}{\Large{\textbf{Sensitivity with the standard approach}}}\\
\cline{1-5} 
$\hat{k}$ & $\partial_{\varsigma}T$ & $\sigma$ & $N_{1\%}$ & $\bar{R}$ \\
\hline 
\hline
\cline{1-5} 
$\hat{k}_1$ & -1.915E-03 & 5.910E-05 & 95E+03 & 194.6\\
\cline{1-5} 
$\hat{k}_5$ & -1.8357E-03 & 6.347E-05 & 120E+03 & 38.7\\
\cline{1-5} 
$\hat{k}_{10}$ & -1.868E-03 & 1.008E-04 & 304E+03 & 20.5 \\
\cline{1-5} 
$\hat{k}_{100}$ & -1.852E-03 & 2.250E-04 & 1516E+03 & 3.4\\
\cline{1-5}
$\hat{k}_{1000}$ & -1.491E-03 & 5.083E-04 & - & 1.8  \\
\hline
\end{tabular}
\label{table2}
                         }\\
    \subfloat[
Evaluation of sensitivity $\partial_{\varsigma}T$ with Alg.~\ref{alg:AlgoDf2}.]{
\begin{tabular}{|L{1.5cm} || C{2.25cm} | C{2.25cm} | C{2.25cm} | C{2.25cm} | }
\cline{2-5}
%\cellcolor{lightgray}
\multicolumn{1}{c|}{}  & \multicolumn{4}{c |}{ \Large{\textbf{Sensitivity with the alternative approach}}} \\
\cline{1-5} 
$\hat{k}$ & $\partial_{\varsigma}T$ & $\sigma$ & $N_{1\%}$ & $\bar{R}$\\
\hline 
\hline
\cline{1-5} 
$\hat{k}_1$  & -1.801E-03 & 2.836E-05 & 24E+03 & 288.4 \\
\cline{1-5} 
$\hat{k}_5$ & -1.838E-03 & 2.745E-05 & 22E+03  & 63.4\\
\cline{1-5} 
$\hat{k}_{10}$ & -1.819E-03 & 2.720E-05 & 22E+03 & 37.7 \\
\cline{1-5} 
$\hat{k}_{100}$ & -1.822E-03 & 2.703E-05 & 22E+03 & 83.6 \\
\cline{1-5}
$\hat{k}_{1000}$ & -1.807E-03 & 2.622E-05 & 20E+03 & 792.4\\
\hline
\end{tabular}
\label{table3}
                         }
\caption{
Simulation results for the \emph{heterogeneous-slab} configuration: evaluation of the transmissivity $T$ of a non diffusive heterogeneous column of length $L$ where the absorption coefficient profile is such that $k_a(\bm{\vec{x}},\varsigma)L=\left(\varsigma-\gamma \right)\frac{\mathrm{atan}(-\alpha\left(\bm{\vec{x}}-\beta\right)+\frac{\pi}{2})}{\pi/2}+\gamma$ with $\varsigma=200, \gamma=0.0, \alpha=1000, \beta=0.0005L$. $T$ and $\partial_\varsigma T$ are evaluated using $N=10000$ samples. The exact solution can be obtained by solving the radiative transfer equation analytically: $T_{exact}=0.326$ and $\partial_{\varsigma} T_{exact}=-1.827E-03$. $N_{1\%}$ is the number of paths that need to be sampled in order to reach $\sigma=\frac{T_{exact}}{100}$ for transmissivity or $\sigma=\frac{\partial_kT_{exact}}{100}$ for sensitivity. $\bar{C}$ is the average number of null-collision per path and $\bar{R}$ the average number of random generation per path.
Table~\ref{table1} confirms that the computation time decreases when $\hat{k}$ gets closer to $k_a$: the average number of random generation per path decreases when $\hat{k}$ is well adjusted because there are less null-collisions.
Table~\ref{table2} highlights the fact that the better $\hat{k}$ is adjusted, the greater is the standard deviation of the sensitivity estimator.
With Table~\ref{table3} we see that with the alternative approach the standard deviation of the sensitivity estimator is independent of $\hat{k}$. The fact that $\bar{R}$ increases when adjusting $\hat{k}$ in Table~\ref{table3} is associated to the repeated sampling of $L$ each time a mesh is crossed. This could be replaced with only one sample along the complete path but using importance sampling on the basis of the information carried by the acceleration grid (see \ref{appendix:additional_sampling}).
}
\label{Table}
  \end{center}
\end{figure}
\FloatBarrier

\begin{figure}[h]
  \begin{center}
\begin{tabular}{|L{1.5cm} | L{1.5cm} || C{1.6cm} | C{1.5cm} | C{1.5cm} | C{1.5cm} | C{1.5cm} | C{1.5cm} | C{1.5cm} |}
\cline{3-9}
%\cellcolor{lightgray}
\multicolumn{2}{c|}{} & \multicolumn{7}{c |}{\Large{\textbf{Alternative approach}}}\\
\cline{1-9} 
$k_{a,max}L$ & $k_{s,max}L$ & $\mathcal{A}$ & $\sigma_\mathcal{A}$ &$\partial_{k_{a,max}}\mathcal{A}$ & $\sigma_{\partial_{k_{a,max}}\mathcal{A}}$ & $\partial_{k_{s,max}}\mathcal{A}$ & $\sigma_{\partial_{k_{s,max}}\mathcal{A}}$ &  $t(s)$   \\
\hline 
\hline
\cline{1-9} 
   0.1  &   0.1    & -0.483582    &    8.58E-05    &  0.160386    &    8.37E-04    &  0.002091    &    1.72E-04    &  0.79   \\  \cline{1-9}
   0.1  &   1.0    & -0.481834    &    9.00E-05    &  0.176627    &    8.74E-04    &  0.001992    &    5.89E-05    &  1.28   \\  \cline{1-9}
   0.1  &   3.0    & -0.477867    &    9.92E-05    &  0.212956    &    9.53E-04    &  0.002097    &    4.17E-05    &  2.69   \\  \cline{1-9}
   0.1  &  10.0    & -0.463074    &    1.26E-04    &  0.342196    &    1.17E-03    &  0.002190    &    3.82E-05    & 10.91   \\  \cline{1-9}
   1.0  &   0.1    & -0.366069    &    2.09E-04    &  0.106597    &    1.69E-04    &  0.010256    &    4.19E-04    &  0.91   \\  \cline{1-9}
   1.0  &   1.0    & -0.356424    &    2.13E-04    &  0.110277    &    1.70E-04    &  0.010256    &    1.42E-04    &  1.31   \\  \cline{1-9}
   1.0  &   3.0    & -0.335872    &    2.20E-04    &  0.116637    &    1.72E-04    &  0.009890    &    9.37E-05    &  2.31   \\  \cline{1-9}
   1.0  &  10.0    & -0.276546    &    2.28E-04    &  0.125441    &    1.91E-04    &  0.007140    &    6.91E-05    &  7.16   \\  \cline{1-9}
   3.0  &   0.1    & -0.219088    &    2.21E-04    &  0.049387    &    6.51E-05    &  0.011163    &    5.03E-04    &  1.04   \\  \cline{1-9}
   3.0  &   1.0    & -0.209163    &    2.18E-04    &  0.047957    &    6.92E-05    &  0.010262    &    1.63E-04    &  1.28   \\  \cline{1-9}
   3.0  &   3.0    & -0.190149    &    2.10E-04    &  0.044937    &    7.74E-05    &  0.008717    &    9.89E-05    &  1.90   \\  \cline{1-9}
   3.0  &  10.0    & -0.143655    &    1.83E-04    &  0.035626    &    9.88E-05    &  0.005079    &    6.15E-05    &  4.16   \\  \cline{1-9}
  10.0  &   0.1    & -0.071489    &    1.19E-04    &  0.008151    &    3.47E-05    &  0.003764    &    4.13E-04    &  1.00   \\  \cline{1-9}
  10.0  &   1.0    & -0.068583    &    1.15E-04    &  0.007778    &    3.54E-05    &  0.002880    &    1.30E-04    &  1.08   \\  \cline{1-9}
  10.0  &   3.0    & -0.063426    &    1.06E-04    &  0.006771    &    3.69E-05    &  0.002478    &    7.59E-05    &  1.27   \\  \cline{1-9}
  10.0  &  10.0    & -0.050710    &    8.50E-05    &  0.004806    &    4.00E-05    &  0.001365    &    4.27E-05    &  1.93   \\  \cline{1-9}
\hline
\end{tabular}%
\end{center}      
\caption{The \emph{heterogeneous-cube} configuration: $f_{max}^{eq}=1$, $g=0$, $\bm{\vec{x_0}}=(0,0,0)$, and $N=10^6$ samples. Evaluation of $\mathcal{A}=\frac{A}{4\pi k_a(\bm{\vec{x_0}}) f_{max}^{eq}}$ (the stationary net-power density) and its sensitivities $\partial_{k_{a,max}}\mathcal{A}$ and $\partial_{k_{s,max}}\mathcal{A}$ using the alternative approach of Alg.~\ref{alg:AlgoDf2} for $\hat{k}_1=k_{a,max}+k_{s,max}$. We check here that the alternative approach recovers the results of Table~1 in \citep{galtier2013} with $\zeta=1$ (the extinction criterion defined in \citep{galtier2016}). The computation times displayed in the last column correspond to an Intel Core i5 - 2.5Ghz without parallelization.}
\label{Table_cube}
\end{figure}
\FloatBarrier
\begin{figure}[h]
%\begin{adjustwidth}{-2cm}{0cm}
\begin{center}
\subfloat[$\zeta=1$]{
\begin{tabular}{|L{1.5cm} || C{1.5cm} | C{1.5cm}|C{1.5cm} | C{1.5cm} || C{1.5cm} | C{1.5cm} | C{1.5cm} | C{1.5cm} | }
\cline{2-9}
\cline{2-9}
\multicolumn{1}{c|}{}  & \multicolumn{4}{c |}{\Large{\textbf{Standard approach}}}& \multicolumn{4}{c |}{\Large{\textbf{Alternative approach}}}\\
\cline{2-9} 
\multicolumn{1}{c|}{} & $\partial_{k_{a,max}}\mathcal{A}$ & $\sigma_{\partial_{k_{a,max}}\mathcal{A}}$ &$\partial_{k_{s,max}}\mathcal{A}$ & $\sigma_{\partial_{k_{s,max}}\mathcal{A}}$ & $\partial_{k_{a,max}}\mathcal{A}$ & $\sigma_{\partial_{k_{a,max}}\mathcal{A}}$ &$\partial_{k_{s,max}}\mathcal{A}$ & $\sigma_{\partial_{k_{s,max}}\mathcal{A}}$  \\
\hline 
\hline
\cline{1-9} 
$\hat{k}_{1}$ &  0.110196    &   1.97E-04    &  0.010612    &   1.75E-04     &  0.110236    &   1.69E-04    &  0.010652    &   1.43E-04     \\ \hline
$\hat{k}_{2^3}$ &    0.110598    &   2.25E-04    &  0.011006    &   2.06E-04 &   0.110258    &   1.69E-04    &  0.010666   &   1.44E-04     \\
\cline{1-9} 
$\hat{k}_{10^3}$      &   0.110202    &   3.95E-04    &  0.010506    &   3.85E-04&  0.110345    &   1.69E-04    &  0.010649    &   1.43E-04  \\
\cline{1-9} 
$\hat{k}_{100^3}$     &   0.109576    &   1.04E-03    &  0.010004    &   1.03E-03  &  0.110104    &   1.69E-04    &  0.010532    &   1.43E-04 \\
\cline{1-9} 
$\hat{k}_{1000^3}$     &  0.114101    &   2.72E-03    &  0.014621    &   2.72E-03  &  0.110199    &   1.69E-04    &  0.010719    &   1.43E-04   \\
\hline
\end{tabular}%
\label{table1_cube_a}
  }\\
  \subfloat[$\zeta=0.1$]{
    \begin{tabular}{|L{1.5cm} || C{1.5cm} | C{1.5cm}|C{1.5cm} | C{1.5cm} || C{1.5cm} | C{1.5cm} | C{1.5cm} | C{1.5cm} | }
\cline{2-9}
\cline{2-9}
\multicolumn{1}{c|}{}  & \multicolumn{4}{c |}{\Large{\textbf{Standard approach}}}& \multicolumn{4}{c |}{\Large{\textbf{Alternative approach}}}\\
\cline{2-9} 
\multicolumn{1}{c|}{} & $\partial_{k_{a,max}}\mathcal{A}$ & $\sigma_{\partial_{k_{a,max}}\mathcal{A}}$ &$\partial_{k_{s,max}}\mathcal{A}$ & $\sigma_{\partial_{k_{s,max}}\mathcal{A}}$ & $\partial_{k_{a,max}}\mathcal{A}$ & $\sigma_{\partial_{k_{a,max}}\mathcal{A}}$ &$\partial_{k_{s,max}}\mathcal{A}$ & $\sigma_{\partial_{k_{s,max}}\mathcal{A}}$  \\
  \hline
$\hat{k}_{1}$     &  0.110457    &   8.41E-05    &  0.010622    &   1.02E-04 &  0.110407    &   8.79E-05    &  0.010572    &   8.14E-05    \\ \hline 
$\hat{k}_{2^3}$  &  0.110434    &   1.15E-04    &  0.010648    &   1.21E-04 &  0.110365    &   9.99E-05    &  0.010579    &   7.92E-05     \\ \hline
$\hat{k}_{10^3}$   &  0.110251    &   2.19E-04    &  0.010490    &   2.17E-04   &  0.110393    &   1.17E-04    &  0.010632    &   7.26E-05     \\ \hline
$\hat{k}_{100^3}$    &  0.110180    &   5.92E-04    &  0.010401    &   5.90E-04  &  0.110406    &   1.22E-04    &  0.010627    &   6.99E-05     \\ \hline
$\hat{k}_{1000^3}$  &  0.112980    &   1.82E-03    &  0.013008    &   1.82E-03  &  0.110533    &   1.22E-04    &  0.010560    &   6.95E-05     \\ \hline
\end{tabular}%
\label{table1_cube_b}
  }
  \end{center}
  \caption{The \emph{heterogeneous-cube} configuration for $k_{a,max}L=1$, $k_{s,max}L=1$, $f_{max}^{eq}=1$, $g=0$, $\bm{\vec{x_0}}=(0,0,0)$ and $N=10^6$ samples. $\zeta=1$ in Table~\ref{table1_cube_a} and $\zeta=0.1$ in \ref{table1_cube_b}. Simulations are made for various adjustements of the $\hat{k}$ profile to the true profile of $k_e = k_a + k_s$ (using the $\hat{k}_n$ notation as defined in Fig.~\ref{fig1}). These Tables confirm the conclusion of Fig.~\ref{Table}: the standard deviation of the alternative approach is independent of $\hat{k}$ whereas it was increasing when $\hat{k}$ was getting close to $k_e$ in the standard approach.}
\label{Table1_cube}
%\end{adjustwidth}
\end{figure}
\FloatBarrier
\begin{figure}[p]
\begin{center}
	\includegraphics[page=1,clip,scale=1]{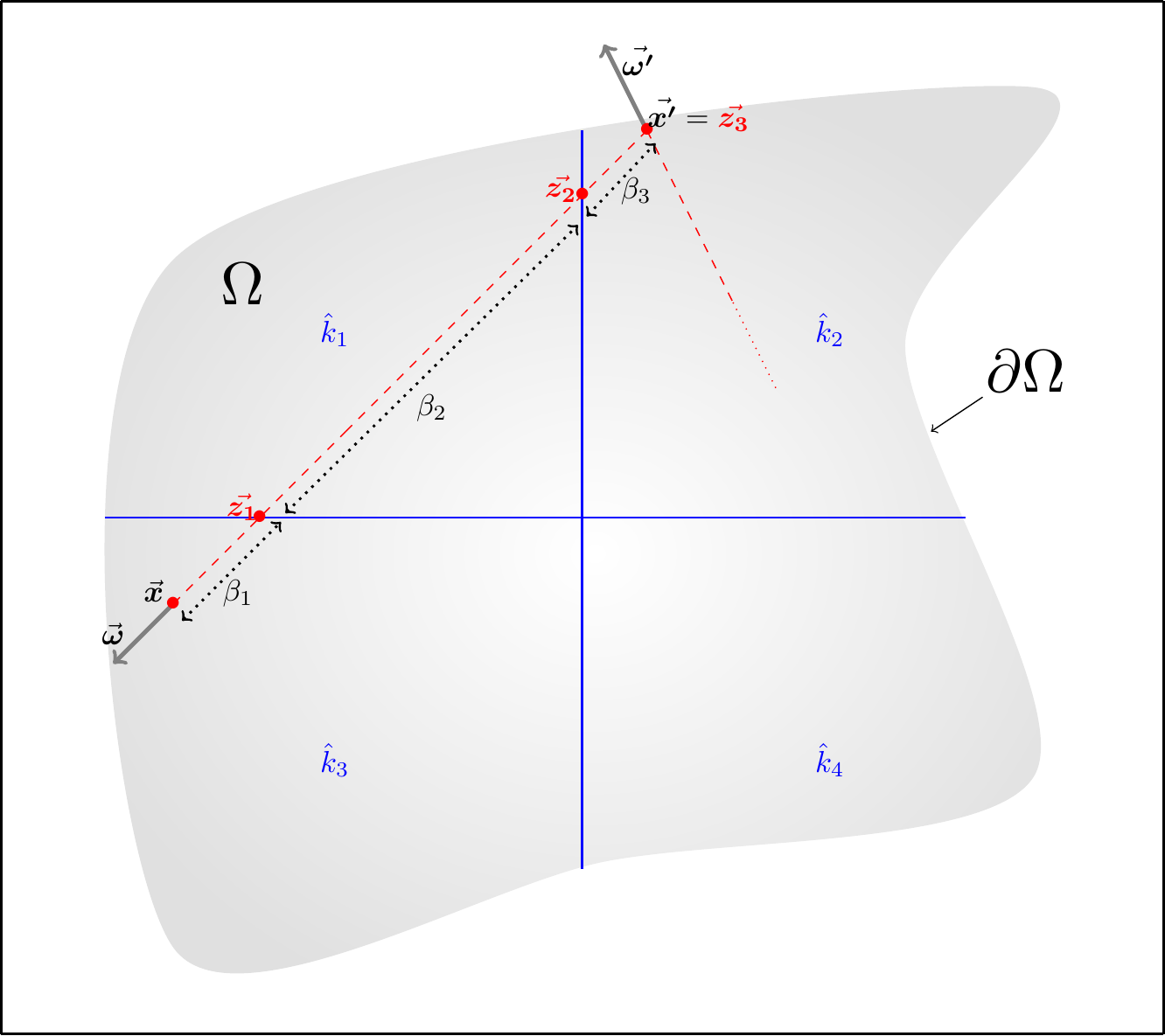}
\end{center}
\caption{
In this 2D sketch, $\Omega$ is partitionned into four areas and $\hat{k}$ is constant in pieces on these areas. We denote $\hat{k}_1$, $\hat{k}_2$, $\hat{k}_3$ and $\hat{k}_4$ these constants. $\bm{\vec{x}}$ and $\bm{\vec{\omega}}$ are the location and direction at the start of the backward tracking of the path. $\bm{\vec{x}'}=\vec{\bm{z_3}}$ is the location of a scattering event (from $\bm{\vec{\omega}'}$ to $\bm{\vec{\omega}}$). $\vec{\bm{z_1}}$ and $\vec{\bm{z_2}}$ are grid-collisions. $\beta_1=||\vec{\bm{z_1}}-\vec{\bm{\vec{x}}}||$, $\beta_2=||\vec{\bm{z_2}}-\vec{\bm{z_1}}||$ and $\beta_3=||\vec{\bm{z_3}}-\vec{\bm{z_2}}||$.}
\label{fig_col_grid}
\end{figure}
\FloatBarrier

%\bibliographystyle{jeb.bst}
%\bibliography{bib_sensib.bib}

\section*{Acknowledgments}
We acknowledge support from the Agence Nationale de la Recherche (ANR, grant HIGH-TUNE ANR-16-CE01-0010, http://www.umr-cnrm.fr/high-tune), from the french Programme National de Teledetection Spatiale (PNTS-2016-05), from Region Occitanie (Projet CLE-2016 EDStar) and from the French Minister of Higher Education, Research and Innovation for the PhD scholarship of the first author.

\section*{References}

\bibliography{mybibfile}

\end{document}